\documentclass{llncs}
\usepackage{pstricks}
\usepackage{xspace}
\usepackage{graphicx}

\usepackage{bold-extra}
\usepackage{listings}

\usepackage{algorithm}
\usepackage{algorithmicx}
\usepackage{algpseudocode}

\usepackage{fancybox}

\makeatletter
\newenvironment{centeredbox}{%
\begin{Sbox}}{
\end{Sbox}\centerline{\parbox{\wd\@Sbox}{\TheSbox}}}
\makeatother

\lstdefinelanguage{JavaScript}{
  keywords={break, case, catch, continue, debugger, default, delete, do, else, finally, for, function, if, in, instanceof, new, return, switch, this, throw, try, typeof, var, void, while, with},
  morecomment=[l]{//},
  morecomment=[s]{/*}{*/},
  morestring=[b]',
  morestring=[b]",
  sensitive=true
}

\begin{document}

\title{Removing Dynamic Type Tests with Context-Driven Basic Block Versioning}
\author{Maxime Chevalier-Boisvert \and Marc Feeley} 
\institute{DIRO, Universit\'e de Montr\'eal, Montreal, QC, Canada
	\email{\{chevalma,feeley\}@iro.umontreal.ca}
}

\maketitle

\begin{abstract}

%

Dynamic typing is an important feature of dynamic programming
languages.  Primitive operators such as those for performing
arithmetic and comparisons typically operate on a wide variety of
input value types, and as such, must internally implement some form of
dynamic type dispatch and type checking.  Removing such type tests is
important for an efficient implementation.

In this paper, we examine the effectiveness of a novel approach to
reducing the number of dynamically executed type tests called {\em
  context-driven basic block versioning}. This simple technique clones
and specializes basic blocks in such a way as to allow the compiler to
accumulate type information while machine code is generated, without a
separate type analysis pass.  The accumulated information allows the
removal of some redundant type tests, particularly in
performance-critical paths.

We have implemented intraprocedural context-driven basic block
versioning in a JavaScript JIT compiler. For comparison, we have also
implemented a classical flow-based type analysis operating on the same
concrete types. Our results show that basic block versioning performs
better on most benchmarks and removes a large fraction of type tests
at the expense of a moderate code size increase.  We believe that this
technique offers a good tradeoff between implementation complexity
and performance, and is suitable for integration in production JIT
compilers.

\end{abstract}

\section{Introduction}\label{sec:intro}

%


Dynamic programming languages make heavy use of late binding in their
semantics.  In essence this means doing at run time what can be done
before run time in other programming languages, for example type
checking, type dispatch, function redefinition, code linking, program
evaluation (e.g.\ {\tt eval}), and compilation (e.g.\ JIT
compilation).  In dynamic programming languages such as JavaScript,
Python, Ruby and Scheme, there are no type annotations on variables
and types are instead associated with values.  Primitive operations,
such as {\tt +}, must verify that the operand values are of an
acceptable type (type checking) and must use the types of the values
to select the operation, such as integer addition, floating point
addition, or string concatenation (type dispatching).  We will use the
generic term {\it type test} to mean a run time operation that
determines if a value belongs to a given type.  Type checking and
dispatching are built with type tests.

VMs for dynamic programming languages must tackle the run time
overhead caused by the dynamic features to achieve an efficient
execution.  Clever type representation and runtime system
organization can help reduce the cost of the dynamic features.  In
this paper we focus on reducing the number of type tests
executed, which is a complementary approach.


Static type analyses which infer a type for each variable can help
remove and in some cases eliminate type test cost.  However, such
analyses are of limited applicability in dynamic languages because of
the run time cost and the presence of generic operators.  A whole
program analysis provides good precision, compared to a more local
analysis, but it is time consuming, which is an issue when compilation
is done during program execution.  Moreover, the results are generally
invalidated when functions are redefined and code is dynamically
loaded or evaluated, requiring a new program analysis.  This often
means that analysis precision must be traded for speed.
Intraprocedural analyses are a good compromise when such dynamic
features are used often, the program is large, the running time is
short or a simple VM design is desired.
The complexity of the type hierarchy for the numerical types may
negatively impact the precision of the type analysis due to its
conservative nature.  To implement the numerical types of the language
or for performance, a VM may use several concrete types for numbers
(e.g.\ fixed precision integers, infinite precision integers, floating
point numbers, complex numbers, etc).  The VM automatically converts
from one representation to another when an operator receives
mixed-representation operands or limit cases are encountered (e.g.\
overflows).  Variables containing numbers will often have to be
assigned a type which is the union of some concrete numerical types
(e.g.\ {\tt int} $\cup$ {\tt float}) if they store the result of
an arithmetic operator.  This means that a type dispatch will
have to be executed when this variable is used as an operand of
another arithmetic operator.  This is an important issue due to the
frequent use of arithmetic in typical programs (for example an
innocuous looking {\tt i++} in a loop will typically require
a type dispatch and overflow check).


We propose a new approach which reduces the number of type tests
by eliminating those that are redundant within each function.
Basic block versioning aims to be a simple and efficient technique
mixing code generation, analysis and optimization.  Section 2 explains
the basic block versioning approach in more details.  An
implementation of our approach in a JavaScript compiler is described
in Section 3 and evaluated in Section 4.  Related work is
presented in Section 5.



\section{Basic Block Versioning}\label{sec:versioning}
\begin{figure}
\begin{center}
\begin{minipage}{2in}
\begin{lstlisting}[language=javascript]
function f(n) {
    if (n<0) n = n+1;
    return n+1;
}
\end{lstlisting}
\end{minipage}
\begin{minipage}{2in}
\includegraphics[scale=0.39]{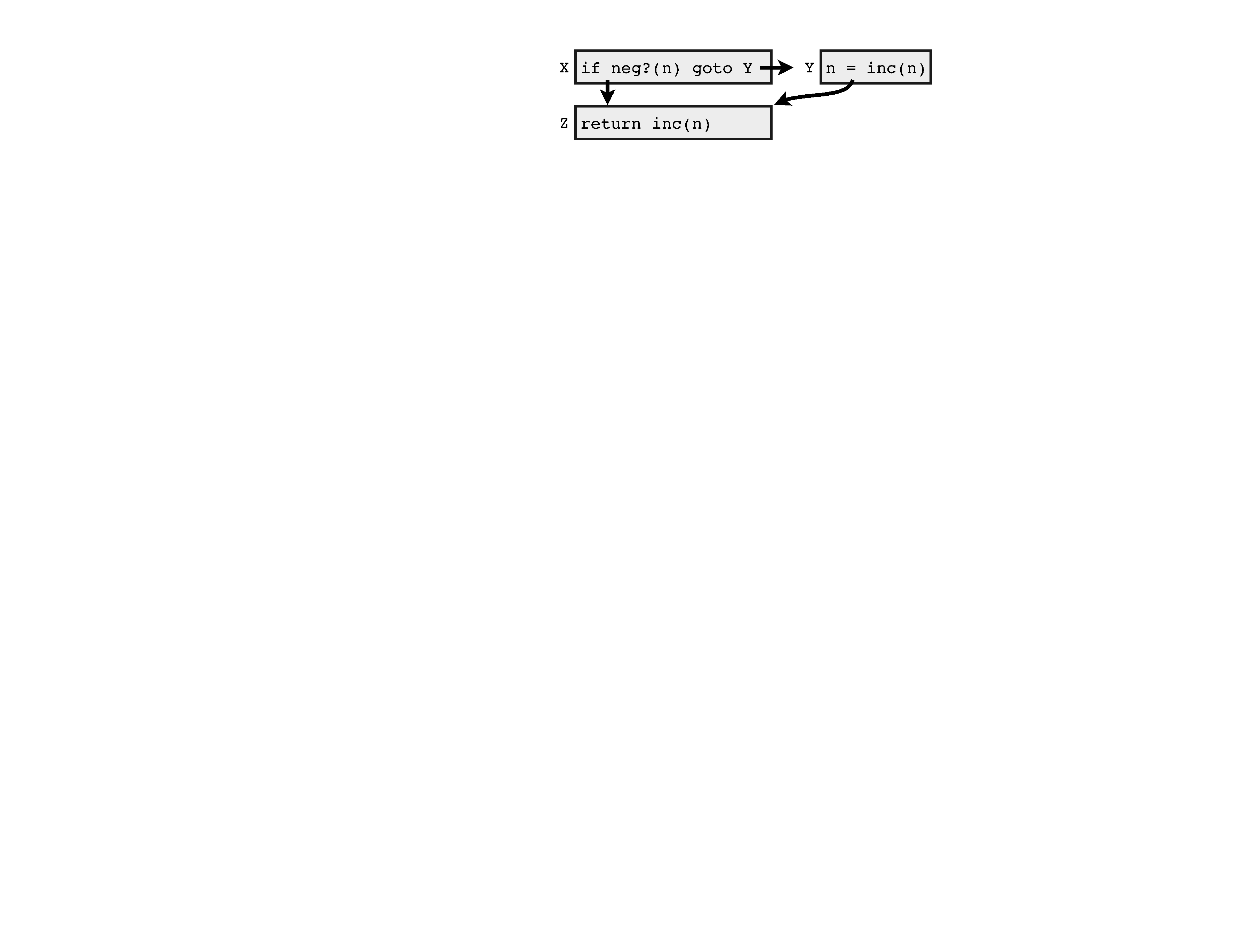}
\end{minipage}
\end{center}
\vspace*{-5ex}
\caption{Definition for function {\tt f} and the corresponding high-level control flow graph.\label{fig:cfg-example}}
\end{figure}

\begin{figure}
\begin{center}
\includegraphics[scale=0.39]{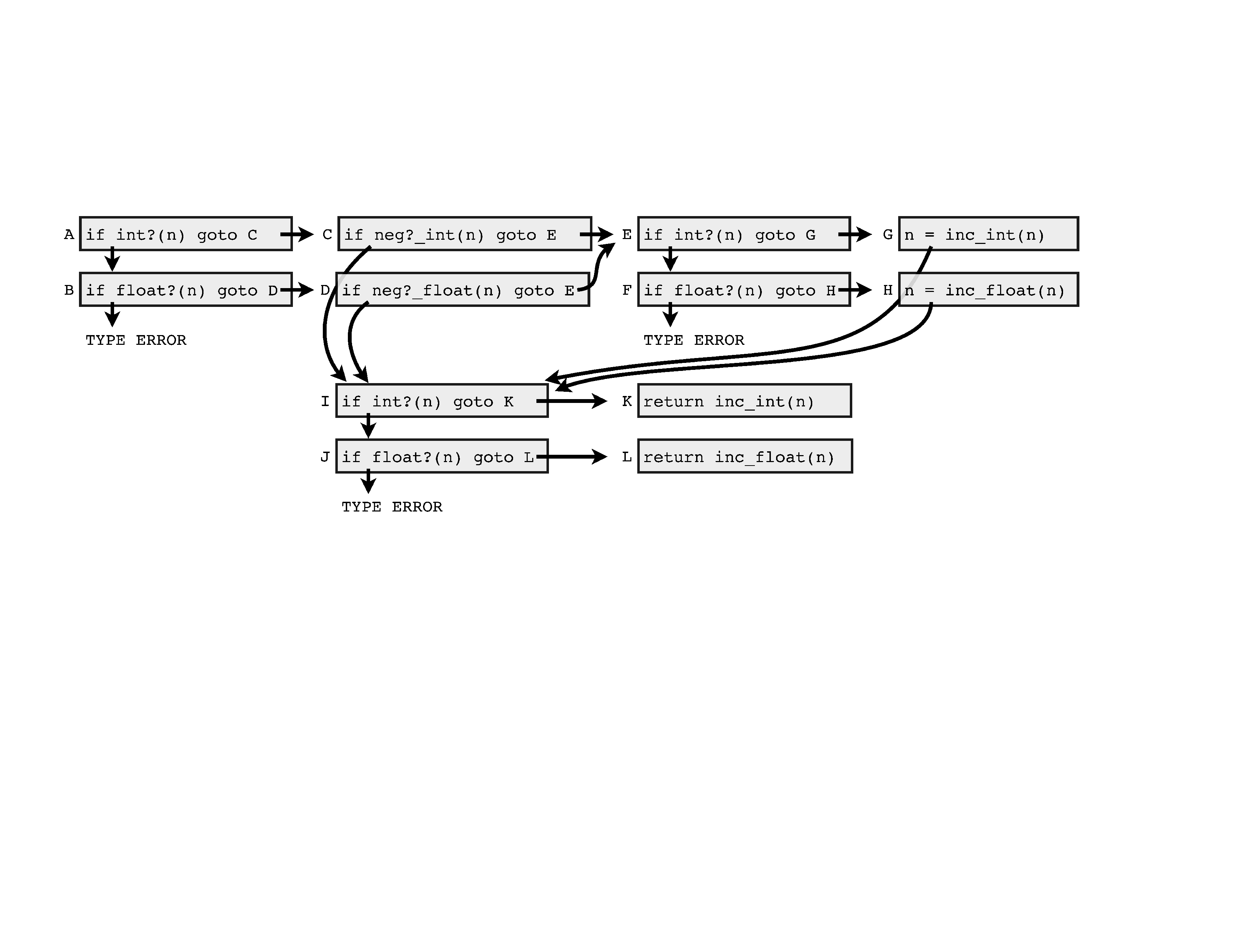}
\end{center}
\vspace*{-5ex}
\caption{Control flow graph after the inlining of the primitive operators {\tt neg?} and {\tt inc}.\label{fig:ll-cfg}}
\end{figure}



The basic block versioning approach generates one or more versions of
each live basic block based on type information derived from the type
tests executed by the code.  The type analysis and code generation are
performed together, generating on-demand new versions of blocks
specialized to the typing context of predecessor blocks.

An important difference between this approach and traditional type
analyses is that basic block versioning does not compute a fixed-point
on types, but rather computes a fixed-point on the generation of new
block versions, each associated with a configuration of incoming
types. Values which have different types at the same program point are
handled more precisely with basic block versioning.  In a traditional
type analysis the union of the possible types would be assigned to the
value, causing the analysis to be conservative.  With basic block
versioning, distinct basic blocks will be created for each
type tested previously, allowing a more precise tracking of types.
Because versions are created on demand, only versions for the relevant
type combinations are created.

To illustrate this approach we will use a simple example in a
hypothetical dynamically typed language similar only in syntax to
JavaScript.  Consider the function {\tt f} whose definition and
corresponding high-level control flow graph are shown in
Figure~\ref{fig:cfg-example}.  Lets assume that there are only two
concrete types for numbers: {\tt int}, a fixed precision integer, and
{\tt float}, a floating point number.\footnote{Note that JavaScript
has a single type for numbers, which corresponds to IEEE 64-bit
floating point numbers, but an implementation of JavaScript could
implement numbers with these two concrete types to benefit from the
performance of integer arithmetic for integer loop iteration variables
and array indexing.}  The value of parameter {\tt n} must be one of
these two types, otherwise it is a type error.  The primitive
operations {\tt neg?(n)} and {\tt inc(n)} must include a type dispatch
to select the appropriate operation based on the concrete type of {\tt
n}.  Inlining these primitive operations makes the type tests explicit
as shown in the control flow graph in Figure~\ref{fig:ll-cfg}.  Note
that basic block X has been expanded to basic blocks A-D, while Y has
been expanded to E-H, and Z has been expanded to I-L.  Note that for
simplicity we will assume that the {\tt inc\_int(n)}
operation yields an {\tt int} (i.e.\ there is no overflow check).

Basic block versioning starts compiling basic block A with a context
where value {\tt n} is of an unknown type.  This will
generate the code for the {\tt int?(n)} type test and will schedule the
compilation of a version of block B, called
B$_{\neg{\mbox{\footnotesize\tt int}}}$, where the value {\tt n} is
known to not be an {\tt int} and will schedule the compilation of a
version of block C, called C$_{\mbox{\footnotesize\tt int}}$, where
the value {\tt n} is known to be an {\tt int}.
Our use of subscripts is a purely notational way of keeping track of the
type context information, which only needs to give information
on {\tt n} in this example.
When basic block
C$_{\mbox{\footnotesize\tt int}}$ is compiled, code is generated for
the {\tt neg?\_int(n)} test and this schedules the compilation of
versions of blocks E and I, called E$_{\mbox{\footnotesize\tt int}}$
and I$_{\mbox{\footnotesize\tt int}}$ respectively, where the value
{\tt n} is known to be an {\tt int}.  Note that the compilation of
B$_{{\neg{\mbox{\footnotesize\tt int}}}}$ will cause the compilation
of D$_{\mbox{\footnotesize\tt float}}$, which will also schedule the
compilation of versions of blocks E and I but in a different context,
where the value {\tt n} is known to be a {\tt float} (blocks
E$_{\mbox{\footnotesize\tt float}}$ and I$_{\mbox{\footnotesize\tt
float}}$ respectively).

\begin{figure}
\begin{center}
\hspace*{-2ex}%
\includegraphics[scale=0.39]{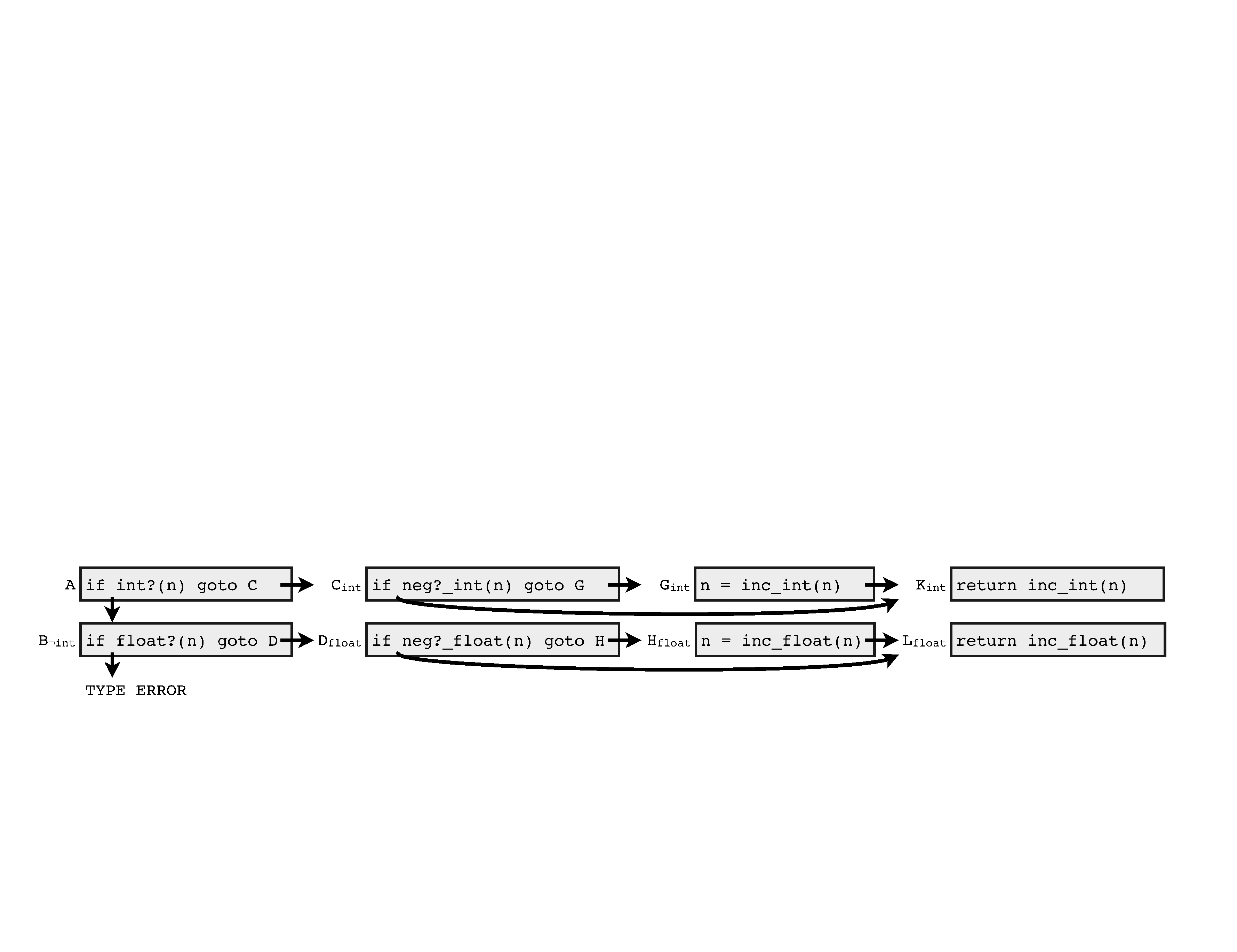}
\end{center}
\vspace*{-5ex}
\caption{Final control flow of function {\tt f} after basic block versioning.\label{fig:final-cfg}}
\end{figure}

The type tests in the four blocks E$_{\mbox{\footnotesize\tt int}}$,
E$_{\mbox{\footnotesize\tt float}}$, I$_{\mbox{\footnotesize\tt int}}$
and I$_{\mbox{\footnotesize\tt float}}$ can be removed and replaced by
direct jumps to the appropriate destination blocks.  For example
E$_{\mbox{\footnotesize\tt int}}$ becomes a direct jump to
G$_{\mbox{\footnotesize\tt int}}$ and E$_{\mbox{\footnotesize\tt
float}}$ becomes a direct jump to F$_{\mbox{\footnotesize\tt float}}$.
Because G$_{\mbox{\footnotesize\tt int}}$ and
H$_{\mbox{\footnotesize\tt float}}$ jump respectively to
I$_{\mbox{\footnotesize\tt int}}$ and I$_{\mbox{\footnotesize\tt
float}}$, the type tests in those blocks are also removed.  Note that
the final generated code implements the same control flow graph as
Figure~\ref{fig:final-cfg}.  Two of the three type dispatch operations
in the original code have been removed.

Let us now consider what would happen if the {\tt inc\_int(n)}
operations detected integer overflow and yielded a {\tt float} result
in that case.  Then the compilation of basic block
G$_{\mbox{\footnotesize\tt int}}$ would schedule two versions of the
successor basic block I: I$_{\mbox{\footnotesize\tt int}}$ for the
case where there is no overflow, and I$_{\mbox{\footnotesize\tt
float}}$ for the case where there is an overflow.  Due to the normal
removal of type tests, when {\tt inc\_int(n)} would overflow, a {\tt
float} would be stored in {\tt n} followed by a direct jump to block
L$_{\mbox{\footnotesize\tt float}}$.  Thus the only change in the
control flow of Figure~\ref{fig:final-cfg} is that block
G$_{\mbox{\footnotesize\tt int}}$ has an edge to
L$_{\mbox{\footnotesize\tt float}}$ in addition to
K$_{\mbox{\footnotesize\tt int}}$.  So here too a single type dispatch
is needed in function {\tt f}.



In theory, the number of possible type configurations in a context
grows combinatorially with the number of live values whose type is
being accounted for and the number of types they can have.  We believe
that a combinatorial explosion is unlikely to be a problem in practice
because typically the number of values live at a given program point is small
and the number of possible types of a value is small.

There are pathological cases where a large
number of block versions are needed to account for all
the possible incoming type combinations. To prevent such
occurrences, a simple approach is to place an arbitrary limit per block
on the number of versions that are compiled. Once this limit
is hit for a given block, a general version of the block will be
compiled, which makes no assumptions about incoming types,
i.e.\ all values are of unknown type.

If more versions of a block would be required than the block version
limit allows, it is advantageous to compile the versions which will be
executed most often.  This can be done by monitoring the frequency of
execution of each basic block (with a counter per block) prior to the
JIT compilation.  Generating linear machine code sequences along hot
paths first has the beneficial effect that it will tend to prioritize
the compilation of block versions for type combinations that occur
more frequently at run time.  This strategy is used in our
experiments.

An indirect benefit of our basic block versioning approach is that it
automatically unrolls some of the first iterations of loops in such a
way that type tests are hoisted out of loop bodies. For example, if
variables of unknown type are used unconditionnaly in a loop, their
type will be tested inside the first iteration of the loop. Once this
test is performed in the first loop iteration, the type information
gained will allow the loop body to avoid the redundant type tests
for the remaining iterations.

\section{Implementation in Higgs}\label{sec:implementation}

We have implemented basic block versioning inside a JavaScript virtual machine
called Higgs. This virtual machine comprises an interpreter and a
JIT compiler targeted at x86-64 POSIX platforms. The current implementation of
Higgs supports most of the ECMAScript 5 specification~\cite{js_spec},
with the exception of the {\tt with} statement, property attributes
and getter-setter properties. Its runtime and standard libraries are
self-hosted, written in an extended dialect of ECMAScript with low-level
primitives. These low-level primitives are special IR instructions which allow
us to express type tests as well as integer and floating point machine
instructions in the implementation language. 

In Higgs, the interpreter is used for profiling, and as a default, unoptimized
mode of execution. Functions are parsed into an abstract syntax tree, and
lazily compiled to an Static Single Assignment (SSA) Intermediate
Representation (IR) when they are first called. The interpreter then executes
code in SSA form directly. As code is executed by the interpreter, counters on
basic blocks are incremented every time a given block is executed. Frequency
counts for each potential callee are also collected at call sites.

The JIT compiler is triggered when the execution count for a function entry
block or loop header reaches a fixed threshold (currently set to 800). Callees
are first aggressively inlined into the function to be compiled. This is done
by substituting the IR of callees at call sites. Calls are currently inlined
only if profiling data indicates that they are monomorphic, and the callee is
30 basic blocks or less, which enables inlining of most runtime primitives.
Call sites belonging to blocks with higher execution frequencies are
prioritized for inlining. Once inlining is complete, the fused IR containing
inlined callees is then optimized using simple subgraph substitution patterns
before machine code generation proceeds.

\begin{algorithm}
\caption{Code generation with basic block versioning}\label{alg:code_gen}
\begin{algorithmic}[1]

\Procedure{genFun}{$assembler$, $function$}

    \State $workList \leftarrow \emptyset$ \Comment{Stack of block versions to be compiled}

    \State $versionList \leftarrow \emptyset$ \Comment{List of existing block versions}

    \State getLabel(function.entryBlock, $\emptyset$, workList, versionList) \Comment{Begin compilation}

    \While{workList not empty}
        \State block, ctx, label $\gets$ workList.pop()

        \State assembler.addLabel(label) \Comment{Insert the label for this block}

        \If{block.execCount is 0}
            \State genStub(assembler, block, ctx, label)
        \Else
            \For{instr in block.instrs} \Comment{Generate code for each instruction}
                \State genInstr(assembler, instr, ctx, workList, versionList);
            \EndFor
        \EndIf
    \EndWhile

\EndProcedure

\Procedure{getLabel}{$block$, $ctx$, $workList$, $versionList$}

    \If{numVersions(block) $\ge$ maxvers} \Comment{If the version limit for this block was reached}
        \State bestMatch $\gets$ findBestMatch(block, ctx, versionList);
        \If{bestMatch $\neq$ null} \Comment{If a compatible match was found}
            \State \Return bestMatch
        \Else
            \State ctx $\gets$ $\emptyset$ \Comment{Make a generic version accepting all incoming contexts}
        \EndIf
    \EndIf

    \State label $\gets$ newLabel();    
    \State workList.push($\langle block, ctx, label \rangle$); \Comment{Queue the new version to be compiled}
    \State versionList.append($\langle block, ctx, label \rangle$); \Comment{Add the new block version to the list}
    \State \Return label

\EndProcedure

\algstore{bbv}
\end{algorithmic}
\end{algorithm}

\begin{algorithm}
\caption{Code generation with basic block versioning}\label{alg:bbv_gen}
\begin{algorithmic}[1]
\algrestore{bbv}

\Procedure{AddInt32.genInstr}{$assembler$, $instr$, $ctx$, $workList$, $versionList$}
    \State assembler.addInt32(instr.getArgs()) \Comment{Generate the add machine instruction}
    \State ctx.setOutType(instr, int32) \Comment{The output type of AddInt32 is always int32. If an overflow occurs, the result is recomputed using AddFloat64}
\EndProcedure

\Procedure{IsInt32.genInstr}{$assembler$, $instr$, $ctx$, $workList$, $versionList$}
    \State argType $\leftarrow$ ctx.getType(instr.getArg(0))
    \If{argType is int32}
        \State ctx.setOutType(instr, true)
    \ElsIf{argType $\neq$ $\top$}
        \State ctx.setOutType(instr, false)
    \Else
        \State assembler.isInt32(instr.getArgs()) \Comment{Generate code for the type test}
        \State ctx.setOutType(instr, const)
    \EndIf
\EndProcedure

\Procedure{Jump.genInstr}{$assembler$, $instr$, $ctx$, $workList$, $versionList$}
    \State label $\gets$ getLabel(instr.target, ctx, workList, versionList)
    \State assembler.jump(label)
\EndProcedure

\Procedure{IfTrue.genInstr}{$assembler$, $instr$, $ctx$, $workList$, $versionList$}

    \State arg $\gets$ instr.getArg(0)
    \State argType $\gets$ ctx.getType(arg)

    \State trueCtx $\gets$ ctx.copy() \Comment{New context for the true branch}

    \If{arg instanceof IsInt32}
        \State trueCtx.setType(arg.getArg(0), int32)
    \EndIf

    \If{argType is true}
        \State trueLabel $\gets$ getLabel(instr.trueTarget, trueCtx, workList, versionList)
        \State assembler.jump(trueLabel)
    \ElsIf{argType is false}
        \State falseLabel $\gets$ getLabel(instr.falseTarget, ctx, workList, versionList)
        \State assembler.jump(falseLabel)
    \Else

        \State trueLabel $\gets$ getLabel(instr.trueTarget, trueCtx, workList, versionList)
        \State falseLabel $\gets$ getLabel(instr.falseTarget, ctx, workList, versionList)

        \State assembler.compare(arg, true) \Comment{Compare the argument to true}
        \State assembler.jumpIfEqual(trueLabel)
        \State assembler.jump(falseLabel)

    \EndIf

\EndProcedure

\end{algorithmic}
\end{algorithm}

Machine code generation (see Algorithm \ref{alg:code_gen}) begins with the
function's entry block and entry context pair being pushed on top of a stack
which serves as a work list. This stack is used to keep track of block versions
to be compiled, and enable depth-first generation of hot code paths. Code
generation proceeds by repeatedly popping a block and context pair to be
compiled off the stack. If the block to be compiled has an execution count of
0, stub code is generated out of line, which spills live variables, invalidates
the generated machine code for the function and exits to the interpreter.
Otherwise, code is generated by calling code generation methods corresponding
to each IR instruction to be compiled in the current block, in order.

As each IR instruction in a block is compiled, information is both retrieved
from and inserted into the current context. Information retrieved may be used
to optimize the compilation of the current instruction (e.g.\ eliminate type
tests). Instructions will also write their own output type in the context if known.
The last instruction of a block, which must be a branch instruction, may
potentially push additional compilation requests on the work stack. More
specifically, branch instructions can request an assembler label for a version
of a block corresponding to the current context at the branch instruction. If
such a version was already compiled, the label is returned immediately.
Otherwise, a new label is generated, the block and the current context are
pushed on the stack, to be compiled later.

To avoid pathological cases where a large number of versions
could be generated for a given basic block, we limit the number of versions
that may be compiled. This is done with the {\tt maxvers} parameter, which 
specifies how many versions can be compiled for any single block. Once this
limit is hit for a particular block, requests for new versions of this block
will first try to find if an inexact but compatible match for the
incoming context can be found. An existing version is compatible with the
incoming context if the value types assumed by the existing version are the
same as, or supertypes of, those specified in the incoming context. If a compatible
match is found, this match will be returned. If not, a generic version of the
block will be generated, which can accept all incoming type combinations.
When the {\tt maxvers} parameter is set to zero, basic block versioning is
disabled, and only the generic version is generated.

\subsection{Type tags and runtime primitives}

The current version of Higgs segregates values into a few categories based on
type tags~\cite{type_tags}. These categories are: 32-bit integers ({\tt int32}),
64-bit floating point values ({\tt float64}), garbage-collected references
inside the Higgs heap ({\tt refptr}), raw pointers to C objects ({\tt rawptr})
and miscellaneous JavaScript constants ({\tt const}). These type tags form
a simple, first-degree notion of types which we use to drive the basic block
versioning approach. The current implementation of basic block versioning in
Higgs does not differentiate between references to object, arrays and
functions, but instead lumps all of these under the reference pointer
category. We do, however, distinguish between the boolean {\tt true} and
{\tt false} constants to enable the propagation of type test results.

We believe that this choice of a simple type representation is a worthwhile
way to investigate the effectiveness and potential of basic block versioning.
Higgs implements JavaScript operators as runtime library functions written in
an extended dialect of JavaScript, and most of these functions use type tags
to do dynamic dispatch. As such, eliminating this first level of type tests is
crucial to improving the performance of the system as a whole. Extending the
system to use a more precise representation of types is part of future work.

\begin{figure}
\begin{centeredbox}
\begin{lstlisting}[language=javascript]
function $rt_add(x, y) {
    if ($ir_is_i32(x)) { // If x is integer
        if ($ir_is_i32(y)) {
            if (var r = $ir_add_i32_ovf(x, y))
                return r;
            else // Handle the overflow case
                return $ir_add_f64($ir_i32_to_f64(x),
                                   $ir_i32_to_f64(y));
        } else if ($ir_is_f64(y))
            return $ir_add_f64($ir_i32_to_f64(x), y);
    } else if ($ir_is_f64(x)) { // If x is floating point
        if ($ir_is_i32(y))
            return $ir_add_f64(x, $ir_i32_to_f64(y));
        else if ($ir_is_f64(y))
            return $ir_add_f64(x, y);
    }

    // Evaluate arguments as strings and concatenate them
    return $rt_strcat($rt_toString(x), $rt_toString(y));
}
\end{lstlisting}
\end{centeredbox}
\caption{Implementation of the {\tt +} operator\label{fig:rt_add}}
\end{figure}

Figure \ref{fig:rt_add} illustrates the implementation of the primitive {\tt +}
operator. As can be seen, this function makes extensive use of
low-level type test primitives such as {\tt \$ir\_is\_i32} and
{\tt \$ir\_is\_f64} to implement dynamic dispatch based on the type tags
of the input arguments. All other arithmetic and comparison primitives 
implement a similar dispatch mechanism.

\subsection{Flow-based representation analysis}

To provide a point of comparison and contrast the capabilities of basic block
versioning with that of more traditional type analysis approaches, we have
implemented a forward flow-based representation analysis which computes a
fixed-point on the types of SSA values. The analysis is an adaptation of
Wegbreit's algorithm as described in~\cite{sccp}. It is an intraprocedural
constant propagation analysis which propagates the types of SSA values in a
flow-sensitive manner. Pseudocode for this analysis and some of its transfer
functions is shown in Appendix \ref{apx:type_prop}.

\begin{figure}[ht!]
\centering
\includegraphics[scale=0.28]{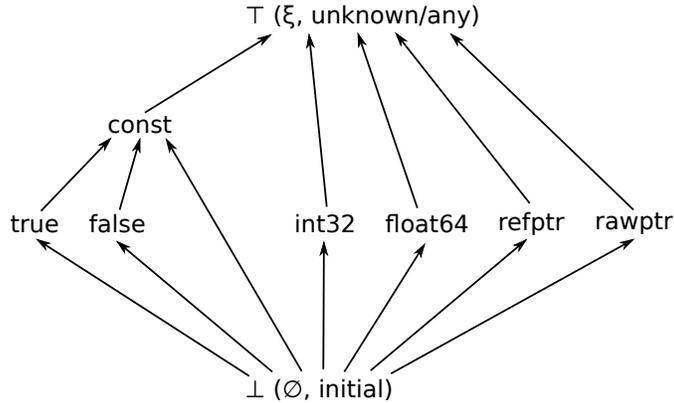}
\caption{Type lattice used by the representation analysis\label{fig:lattice}}
\end{figure}

The representation analysis uses the same type representation (see
Figure \ref{fig:lattice}) as our basic block versioning implementation, and has
similar type analysis capabilities. It is able to gain information from type
tests and forward this information along branches. It is also able to deduce,
in some cases, that specific branches will not be executed and ignore the
effects of code that was determined dead.

We have also extended the flow-based algorithm to ignore basic blocks which
are unexecuted (have an execution count of 0) at analysis time. This allows
the analysis to ignore some code paths not executed up to now, which is useful in
some cases, since primitive language operators often have multiple paths which
can result in different output types. If presumed dead blocks turn out to be
executed later, analysis results and associated compiled code will be
invalidated at run time. This was done to make the analysis more competitive
with basic block versioning which by construction ignores stubbed blocks, for
which no compiled code was generated.

\subsection{Limitations}

There are a few important limitations to the current implementation of
basic block versioning in Higgs. We do not, at this point, track the types of
object properties. Global variables, which are properties of the global object
in JavaScript, are also untracked. We do not account for
interprocedural flow of type information either. That is, function parameter
and return value types are assumed to be unknown. Finally, the current
implementation of Higgs does not implement any kind of load-store forwarding
optimization. These limitations are nontrivial to tackle due to factors such
as the late-bound nature of JavaScript, the potential presence of the
{\tt eval} construct, dynamic addition and deletion of properties and the
dynamic installation of getter-setter methods on object fields.

The results presented in this paper are entirely based on an intraprocedural
implementation of basic block versioning which accounts only for the types of
local variables and temporaries, in combination with aggressive inlining of
library and method calls. Extending basic block versioning to take object
identity, array and property type information into account constitutes future 
work.

\section{Evaluation}\label{sec:evaluation}
To assess the effectiveness of basic block versioning, we have used a total of
24 benchmarks from the classic SunSpider and Google V8 suites. A handful of
benchmarks from both suites were not included in our tests because the
current Higgs implementation does not yet support them.

\begin{figure}
\begin{center}
\bgroup
\psset{unit=1pt}
\begin{pspicture}(334.559,210.037)
\newgray{gray0}{0.}
\newgray{gray25}{.25}
\newgray{gray50}{.5}
\newgray{gray75}{.75}
\newgray{gray100}{1.}
\newgray{gray90}{.9}
\newgray{gray0}{0}
\psframe[linewidth=1.,linecolor=gray0,fillcolor=gray90,fillstyle=solid](27.5,68.537)(328.5,189.537)
\psframe[linewidth=1.,linecolor=gray0,fillcolor=gray100,fillstyle=solid](27.5,200.537)(36.5,209.537)
\rput[Bl]{0}(38.,201.83700000000002){analysis}
\psframe[linewidth=1.,linecolor=gray0,fillcolor=gray75,fillstyle=solid](87.5,200.537)(96.5,209.537)
\rput[Bl]{0}(98.,201.83700000000002){maxvers=1}
\psframe[linewidth=1.,linecolor=gray0,fillcolor=gray50,fillstyle=solid](147.5,200.537)(156.5,209.537)
\rput[Bl]{0}(158.,201.83700000000002){maxvers=2}
\psframe[linewidth=1.,linecolor=gray0,fillcolor=gray25,fillstyle=solid](207.5,200.537)(216.5,209.537)
\rput[Bl]{0}(218.,201.83700000000002){maxvers=5}
\psframe[linewidth=1.,linecolor=gray0,fillcolor=gray0,fillstyle=solid](267.5,200.537)(276.5,209.537)
\rput[Bl]{0}(278.,201.83700000000002){maxvers=$\infty$}
\psline[linewidth=.5,linecolor=gray0](24.5,69.037)(28.,69.037)
\rput[Br]{0}(21.,66.937){0\%}
\psline[linewidth=.5,linecolor=gray0](24.5,99.037)(28.,99.037)
\psline[linewidth=.5,linecolor=gray0,linestyle=dashed,dash=2 4](28.,99.037)(328.,99.037)
\rput[Br]{0}(21.,96.937){25\%}
\psline[linewidth=.5,linecolor=gray0](24.5,129.037)(28.,129.037)
\psline[linewidth=.5,linecolor=gray0,linestyle=dashed,dash=2 4](28.,129.037)(328.,129.037)
\rput[Br]{0}(21.,126.937){50\%}
\psline[linewidth=.5,linecolor=gray0](24.5,159.037)(28.,159.037)
\psline[linewidth=.5,linecolor=gray0,linestyle=dashed,dash=2 4](28.,159.037)(328.,159.037)
\rput[Br]{0}(21.,156.937){75\%}
\psline[linewidth=.5,linecolor=gray0](24.5,189.037)(28.,189.037)
\rput[Br]{0}(21.,186.937){100\%}
\rput[Bl]{-70.}(40.5,61.037){partial-sums}
\psframe[linewidth=1.,linecolor=gray0,fillcolor=gray100,fillstyle=solid](30.385,68.537)(35.231,174.339)
\psframe[linewidth=1.,linecolor=gray0,fillcolor=gray75,fillstyle=solid](34.231,68.537)(39.077,167.25300000000001)
\psframe[linewidth=1.,linecolor=gray0,fillcolor=gray50,fillstyle=solid](38.077,68.537)(42.923,112.887)
\psframe[linewidth=1.,linecolor=gray0,fillcolor=gray25,fillstyle=solid](41.923,68.537)(46.769,112.024)
\psframe[linewidth=1.,linecolor=gray0,fillcolor=gray0,fillstyle=solid](45.769,68.537)(50.615,112.024)
\rput[Bl]{-70.}(65.5,61.037){v8-crypto}
\psframe[linewidth=1.,linecolor=gray0,fillcolor=gray100,fillstyle=solid](55.385,68.537)(60.231,123.126)
\psframe[linewidth=1.,linecolor=gray0,fillcolor=gray75,fillstyle=solid](59.231,68.537)(64.077,119.876)
\psframe[linewidth=1.,linecolor=gray0,fillcolor=gray50,fillstyle=solid](63.077,68.537)(67.923,119.546)
\psframe[linewidth=1.,linecolor=gray0,fillcolor=gray25,fillstyle=solid](66.923,68.537)(71.769,110.749)
\psframe[linewidth=1.,linecolor=gray0,fillcolor=gray0,fillstyle=solid](70.769,68.537)(75.615,110.749)
\rput[Bl]{-70.}(90.5,61.037){3d-morph}
\psframe[linewidth=1.,linecolor=gray0,fillcolor=gray100,fillstyle=solid](80.385,68.537)(85.23100000000001,177.755)
\psframe[linewidth=1.,linecolor=gray0,fillcolor=gray75,fillstyle=solid](84.23100000000001,68.537)(89.077,177.755)
\psframe[linewidth=1.,linecolor=gray0,fillcolor=gray50,fillstyle=solid](88.077,68.537)(92.923,100.933)
\psframe[linewidth=1.,linecolor=gray0,fillcolor=gray25,fillstyle=solid](91.923,68.537)(96.769,98.58)
\psframe[linewidth=1.,linecolor=gray0,fillcolor=gray0,fillstyle=solid](95.769,68.537)(100.61500000000001,98.58)
\rput[Bl]{-70.}(115.5,61.037){fasta}
\psframe[linewidth=1.,linecolor=gray0,fillcolor=gray100,fillstyle=solid](105.385,68.537)(110.23100000000001,136.923)
\psframe[linewidth=1.,linecolor=gray0,fillcolor=gray75,fillstyle=solid](109.23100000000001,68.537)(114.077,150.78900000000002)
\psframe[linewidth=1.,linecolor=gray0,fillcolor=gray50,fillstyle=solid](113.077,68.537)(117.923,127.886)
\psframe[linewidth=1.,linecolor=gray0,fillcolor=gray25,fillstyle=solid](116.923,68.537)(121.769,122.285)
\psframe[linewidth=1.,linecolor=gray0,fillcolor=gray0,fillstyle=solid](120.769,68.537)(125.61500000000001,122.285)
\rput[Bl]{-70.}(140.5,61.037){bits-in-byte}
\psframe[linewidth=1.,linecolor=gray0,fillcolor=gray100,fillstyle=solid](130.385,68.537)(135.231,87.057)
\psframe[linewidth=1.,linecolor=gray0,fillcolor=gray75,fillstyle=solid](134.231,68.537)(139.077,87.07000000000001)
\psframe[linewidth=1.,linecolor=gray0,fillcolor=gray50,fillstyle=solid](138.077,68.537)(142.923,87.051)
\psframe[linewidth=1.,linecolor=gray0,fillcolor=gray25,fillstyle=solid](141.923,68.537)(146.769,71.798)
\psframe[linewidth=1.,linecolor=gray0,fillcolor=gray0,fillstyle=solid](145.769,68.537)(150.615,71.798)
\rput[Bl]{-70.}(165.5,61.037){crypto-md5}
\psframe[linewidth=1.,linecolor=gray0,fillcolor=gray100,fillstyle=solid](155.385,68.537)(160.231,144.627)
\psframe[linewidth=1.,linecolor=gray0,fillcolor=gray75,fillstyle=solid](159.231,68.537)(164.077,127.99900000000001)
\psframe[linewidth=1.,linecolor=gray0,fillcolor=gray50,fillstyle=solid](163.077,68.537)(167.923,126.74300000000001)
\psframe[linewidth=1.,linecolor=gray0,fillcolor=gray25,fillstyle=solid](166.923,68.537)(171.769,126.337)
\psframe[linewidth=1.,linecolor=gray0,fillcolor=gray0,fillstyle=solid](170.769,68.537)(175.615,126.337)
\rput[Bl]{-70.}(190.5,61.037){deltablue}
\psframe[linewidth=1.,linecolor=gray0,fillcolor=gray100,fillstyle=solid](180.385,68.537)(185.231,152.957)
\psframe[linewidth=1.,linecolor=gray0,fillcolor=gray75,fillstyle=solid](184.231,68.537)(189.077,135.575)
\psframe[linewidth=1.,linecolor=gray0,fillcolor=gray50,fillstyle=solid](188.077,68.537)(192.923,128.92600000000002)
\psframe[linewidth=1.,linecolor=gray0,fillcolor=gray25,fillstyle=solid](191.923,68.537)(196.769,128.62800000000001)
\psframe[linewidth=1.,linecolor=gray0,fillcolor=gray0,fillstyle=solid](195.769,68.537)(200.615,128.627)
\rput[Bl]{-70.}(215.5,61.037){recursive}
\psframe[linewidth=1.,linecolor=gray0,fillcolor=gray100,fillstyle=solid](205.385,68.537)(210.231,189.13)
\psframe[linewidth=1.,linecolor=gray0,fillcolor=gray75,fillstyle=solid](209.231,68.537)(214.077,189.13400000000001)
\psframe[linewidth=1.,linecolor=gray0,fillcolor=gray50,fillstyle=solid](213.077,68.537)(217.923,189.089)
\psframe[linewidth=1.,linecolor=gray0,fillcolor=gray25,fillstyle=solid](216.923,68.537)(221.769,189.08700000000002)
\psframe[linewidth=1.,linecolor=gray0,fillcolor=gray0,fillstyle=solid](220.769,68.537)(225.615,189.08700000000002)
\rput[Bl]{-70.}(240.5,61.037){3d-raytrace}
\psframe[linewidth=1.,linecolor=gray0,fillcolor=gray100,fillstyle=solid](230.385,68.537)(235.231,100.056)
\psframe[linewidth=1.,linecolor=gray0,fillcolor=gray75,fillstyle=solid](234.231,68.537)(239.077,98.257)
\psframe[linewidth=1.,linecolor=gray0,fillcolor=gray50,fillstyle=solid](238.077,68.537)(242.923,96.552)
\psframe[linewidth=1.,linecolor=gray0,fillcolor=gray25,fillstyle=solid](241.923,68.537)(246.769,96.533)
\psframe[linewidth=1.,linecolor=gray0,fillcolor=gray0,fillstyle=solid](245.769,68.537)(250.615,96.526)
\rput[Bl]{-70.}(265.5,61.037){navier-stokes}
\psframe[linewidth=1.,linecolor=gray0,fillcolor=gray100,fillstyle=solid](255.38500000000002,68.537)(260.231,159.135)
\psframe[linewidth=1.,linecolor=gray0,fillcolor=gray75,fillstyle=solid](259.231,68.537)(264.077,177.978)
\psframe[linewidth=1.,linecolor=gray0,fillcolor=gray50,fillstyle=solid](263.077,68.537)(267.923,165.055)
\psframe[linewidth=1.,linecolor=gray0,fillcolor=gray25,fillstyle=solid](266.923,68.537)(271.769,138.24)
\psframe[linewidth=1.,linecolor=gray0,fillcolor=gray0,fillstyle=solid](270.769,68.537)(275.615,138.082)
\rput[Bl]{-70.}(290.5,61.037){nbody}
\psframe[linewidth=1.,linecolor=gray0,fillcolor=gray100,fillstyle=solid](280.385,68.537)(285.231,157.18)
\psframe[linewidth=1.,linecolor=gray0,fillcolor=gray75,fillstyle=solid](284.231,68.537)(289.077,158.68200000000002)
\psframe[linewidth=1.,linecolor=gray0,fillcolor=gray50,fillstyle=solid](288.077,68.537)(292.923,132.29)
\psframe[linewidth=1.,linecolor=gray0,fillcolor=gray25,fillstyle=solid](291.923,68.537)(296.769,130.62)
\psframe[linewidth=1.,linecolor=gray0,fillcolor=gray0,fillstyle=solid](295.769,68.537)(300.615,130.62)
\rput[Bl]{-70.}(315.5,61.037){nsieve-bits}
\psframe[linewidth=1.,linecolor=gray0,fillcolor=gray100,fillstyle=solid](305.385,68.537)(310.231,185.095)
\psframe[linewidth=1.,linecolor=gray0,fillcolor=gray75,fillstyle=solid](309.231,68.537)(314.077,184.42600000000002)
\psframe[linewidth=1.,linecolor=gray0,fillcolor=gray50,fillstyle=solid](313.077,68.537)(317.923,184.373)
\psframe[linewidth=1.,linecolor=gray0,fillcolor=gray25,fillstyle=solid](316.923,68.537)(321.769,184.373)
\psframe[linewidth=1.,linecolor=gray0,fillcolor=gray0,fillstyle=solid](320.769,68.537)(325.615,184.373)
\end{pspicture}
\egroup
\bgroup
\psset{unit=1pt}
\begin{pspicture}(336.2,170.037)
\newgray{gray0}{0.}
\newgray{gray25}{.25}
\newgray{gray50}{.5}
\newgray{gray75}{.75}
\newgray{gray100}{1.}
\newgray{gray90}{.9}
\newgray{gray0}{0}
\psframe[linewidth=1.,linecolor=gray0,fillcolor=gray90,fillstyle=solid](27.5,48.537)(328.5,169.537)
\psline[linewidth=.5,linecolor=gray0](24.5,49.037)(28.,49.037)
\rput[Br]{0}(21.,46.937){0\%}
\psline[linewidth=.5,linecolor=gray0](24.5,79.037)(28.,79.037)
\psline[linewidth=.5,linecolor=gray0,linestyle=dashed,dash=2 4](28.,79.037)(328.,79.037)
\rput[Br]{0}(21.,76.937){25\%}
\psline[linewidth=.5,linecolor=gray0](24.5,109.037)(28.,109.037)
\psline[linewidth=.5,linecolor=gray0,linestyle=dashed,dash=2 4](28.,109.037)(328.,109.037)
\rput[Br]{0}(21.,106.937){50\%}
\psline[linewidth=.5,linecolor=gray0](24.5,139.037)(28.,139.037)
\psline[linewidth=.5,linecolor=gray0,linestyle=dashed,dash=2 4](28.,139.037)(328.,139.037)
\rput[Br]{0}(21.,136.937){75\%}
\psline[linewidth=.5,linecolor=gray0](24.5,169.037)(28.,169.037)
\rput[Br]{0}(21.,166.937){100\%}
\rput[Bl]{-70.}(40.5,41.037){cordic}
\psframe[linewidth=1.,linecolor=gray0,fillcolor=gray100,fillstyle=solid](30.385,48.537)(35.231,166.592)
\psframe[linewidth=1.,linecolor=gray0,fillcolor=gray75,fillstyle=solid](34.231,48.537)(39.077,123.885)
\psframe[linewidth=1.,linecolor=gray0,fillcolor=gray50,fillstyle=solid](38.077,48.537)(42.923,114.96300000000001)
\psframe[linewidth=1.,linecolor=gray0,fillcolor=gray25,fillstyle=solid](41.923,48.537)(46.769,93.514)
\psframe[linewidth=1.,linecolor=gray0,fillcolor=gray0,fillstyle=solid](45.769,48.537)(50.615,93.514)
\rput[Bl]{-70.}(65.5,41.037){spectral-norm}
\psframe[linewidth=1.,linecolor=gray0,fillcolor=gray100,fillstyle=solid](55.385,48.537)(60.231,107.055)
\psframe[linewidth=1.,linecolor=gray0,fillcolor=gray75,fillstyle=solid](59.231,48.537)(64.077,119.787)
\psframe[linewidth=1.,linecolor=gray0,fillcolor=gray50,fillstyle=solid](63.077,48.537)(67.923,110.293)
\psframe[linewidth=1.,linecolor=gray0,fillcolor=gray25,fillstyle=solid](66.923,48.537)(71.769,100.645)
\psframe[linewidth=1.,linecolor=gray0,fillcolor=gray0,fillstyle=solid](70.769,48.537)(75.615,100.644)
\rput[Bl]{-70.}(90.5,41.037){richards}
\psframe[linewidth=1.,linecolor=gray0,fillcolor=gray100,fillstyle=solid](80.385,48.537)(85.23100000000001,128.708)
\psframe[linewidth=1.,linecolor=gray0,fillcolor=gray75,fillstyle=solid](84.23100000000001,48.537)(89.077,120.041)
\psframe[linewidth=1.,linecolor=gray0,fillcolor=gray50,fillstyle=solid](88.077,48.537)(92.923,111.892)
\psframe[linewidth=1.,linecolor=gray0,fillcolor=gray25,fillstyle=solid](91.923,48.537)(96.769,111.891)
\psframe[linewidth=1.,linecolor=gray0,fillcolor=gray0,fillstyle=solid](95.769,48.537)(100.61500000000001,111.89)
\rput[Bl]{-70.}(115.5,41.037){bitwise-and}
\psframe[linewidth=1.,linecolor=gray0,fillcolor=gray100,fillstyle=solid](105.385,48.537)(110.23100000000001,169.3)
\psframe[linewidth=1.,linecolor=gray0,fillcolor=gray75,fillstyle=solid](109.23100000000001,48.537)(114.077,169.302)
\psframe[linewidth=1.,linecolor=gray0,fillcolor=gray50,fillstyle=solid](113.077,48.537)(117.923,169.276)
\psframe[linewidth=1.,linecolor=gray0,fillcolor=gray25,fillstyle=solid](116.923,48.537)(121.769,169.275)
\psframe[linewidth=1.,linecolor=gray0,fillcolor=gray0,fillstyle=solid](120.769,48.537)(125.61500000000001,169.275)
\rput[Bl]{-70.}(140.5,41.037){fannkuch}
\psframe[linewidth=1.,linecolor=gray0,fillcolor=gray100,fillstyle=solid](130.385,48.537)(135.231,158.972)
\psframe[linewidth=1.,linecolor=gray0,fillcolor=gray75,fillstyle=solid](134.231,48.537)(139.077,158.969)
\psframe[linewidth=1.,linecolor=gray0,fillcolor=gray50,fillstyle=solid](138.077,48.537)(142.923,158.964)
\psframe[linewidth=1.,linecolor=gray0,fillcolor=gray25,fillstyle=solid](141.923,48.537)(146.769,158.964)
\psframe[linewidth=1.,linecolor=gray0,fillcolor=gray0,fillstyle=solid](145.769,48.537)(150.615,158.964)
\rput[Bl]{-70.}(165.5,41.037){crypto-sha1}
\psframe[linewidth=1.,linecolor=gray0,fillcolor=gray100,fillstyle=solid](155.385,48.537)(160.231,123.70100000000001)
\psframe[linewidth=1.,linecolor=gray0,fillcolor=gray75,fillstyle=solid](159.231,48.537)(164.077,110.416)
\psframe[linewidth=1.,linecolor=gray0,fillcolor=gray50,fillstyle=solid](163.077,48.537)(167.923,109.336)
\psframe[linewidth=1.,linecolor=gray0,fillcolor=gray25,fillstyle=solid](166.923,48.537)(171.769,108.849)
\psframe[linewidth=1.,linecolor=gray0,fillcolor=gray0,fillstyle=solid](170.769,48.537)(175.615,108.849)
\rput[Bl]{-70.}(190.5,41.037){v8-raytrace}
\psframe[linewidth=1.,linecolor=gray0,fillcolor=gray100,fillstyle=solid](180.385,48.537)(185.231,138.697)
\psframe[linewidth=1.,linecolor=gray0,fillcolor=gray75,fillstyle=solid](184.231,48.537)(189.077,125.811)
\psframe[linewidth=1.,linecolor=gray0,fillcolor=gray50,fillstyle=solid](188.077,48.537)(192.923,107.904)
\psframe[linewidth=1.,linecolor=gray0,fillcolor=gray25,fillstyle=solid](191.923,48.537)(196.769,107.73400000000001)
\psframe[linewidth=1.,linecolor=gray0,fillcolor=gray0,fillstyle=solid](195.769,48.537)(200.615,107.73400000000001)
\rput[Bl]{-70.}(215.5,41.037){3bits-byte}
\psframe[linewidth=1.,linecolor=gray0,fillcolor=gray100,fillstyle=solid](205.385,48.537)(210.231,162.59)
\psframe[linewidth=1.,linecolor=gray0,fillcolor=gray75,fillstyle=solid](209.231,48.537)(214.077,50.238)
\psframe[linewidth=1.,linecolor=gray0,fillcolor=gray50,fillstyle=solid](213.077,48.537)(217.923,50.011)
\psframe[linewidth=1.,linecolor=gray0,fillcolor=gray25,fillstyle=solid](216.923,48.537)(221.769,49.998)
\psframe[linewidth=1.,linecolor=gray0,fillcolor=gray0,fillstyle=solid](220.769,48.537)(225.615,49.997)
\rput[Bl]{-70.}(240.5,41.037){earley-boyer}
\psframe[linewidth=1.,linecolor=gray0,fillcolor=gray100,fillstyle=solid](230.385,48.537)(235.231,128.03)
\psframe[linewidth=1.,linecolor=gray0,fillcolor=gray75,fillstyle=solid](234.231,48.537)(239.077,120.535)
\psframe[linewidth=1.,linecolor=gray0,fillcolor=gray50,fillstyle=solid](238.077,48.537)(242.923,116.19)
\psframe[linewidth=1.,linecolor=gray0,fillcolor=gray25,fillstyle=solid](241.923,48.537)(246.769,115.997)
\psframe[linewidth=1.,linecolor=gray0,fillcolor=gray0,fillstyle=solid](245.769,48.537)(250.615,115.997)
\rput[Bl]{-70.}(265.5,41.037){3d-cube}
\psframe[linewidth=1.,linecolor=gray0,fillcolor=gray100,fillstyle=solid](255.38500000000002,48.537)(260.231,158.82)
\psframe[linewidth=1.,linecolor=gray0,fillcolor=gray75,fillstyle=solid](259.231,48.537)(264.077,90.69800000000001)
\psframe[linewidth=1.,linecolor=gray0,fillcolor=gray50,fillstyle=solid](263.077,48.537)(267.923,89.81)
\psframe[linewidth=1.,linecolor=gray0,fillcolor=gray25,fillstyle=solid](266.923,48.537)(271.769,82.467)
\psframe[linewidth=1.,linecolor=gray0,fillcolor=gray0,fillstyle=solid](270.769,48.537)(275.615,82.101)
\rput[Bl]{-70.}(290.5,41.037){nsieve}
\psframe[linewidth=1.,linecolor=gray0,fillcolor=gray100,fillstyle=solid](280.385,48.537)(285.231,150.41400000000002)
\psframe[linewidth=1.,linecolor=gray0,fillcolor=gray75,fillstyle=solid](284.231,48.537)(289.077,150.416)
\psframe[linewidth=1.,linecolor=gray0,fillcolor=gray50,fillstyle=solid](288.077,48.537)(292.923,150.403)
\psframe[linewidth=1.,linecolor=gray0,fillcolor=gray25,fillstyle=solid](291.923,48.537)(296.769,146.958)
\psframe[linewidth=1.,linecolor=gray0,fillcolor=gray0,fillstyle=solid](295.769,48.537)(300.615,146.958)
\rput[Bl]{-70.}(315.5,41.037){binary-trees}
\psframe[linewidth=1.,linecolor=gray0,fillcolor=gray100,fillstyle=solid](305.385,48.537)(310.231,132.66400000000002)
\psframe[linewidth=1.,linecolor=gray0,fillcolor=gray75,fillstyle=solid](309.231,48.537)(314.077,130.662)
\psframe[linewidth=1.,linecolor=gray0,fillcolor=gray50,fillstyle=solid](313.077,48.537)(317.923,116.33)
\psframe[linewidth=1.,linecolor=gray0,fillcolor=gray25,fillstyle=solid](316.923,48.537)(321.769,116.32900000000001)
\psframe[linewidth=1.,linecolor=gray0,fillcolor=gray0,fillstyle=solid](320.769,48.537)(325.615,116.328)
\end{pspicture}
\egroup
\end{center}
\caption{Counts of dynamic type tests (relative to baseline)\label{fig:test_counts}}
\end{figure}
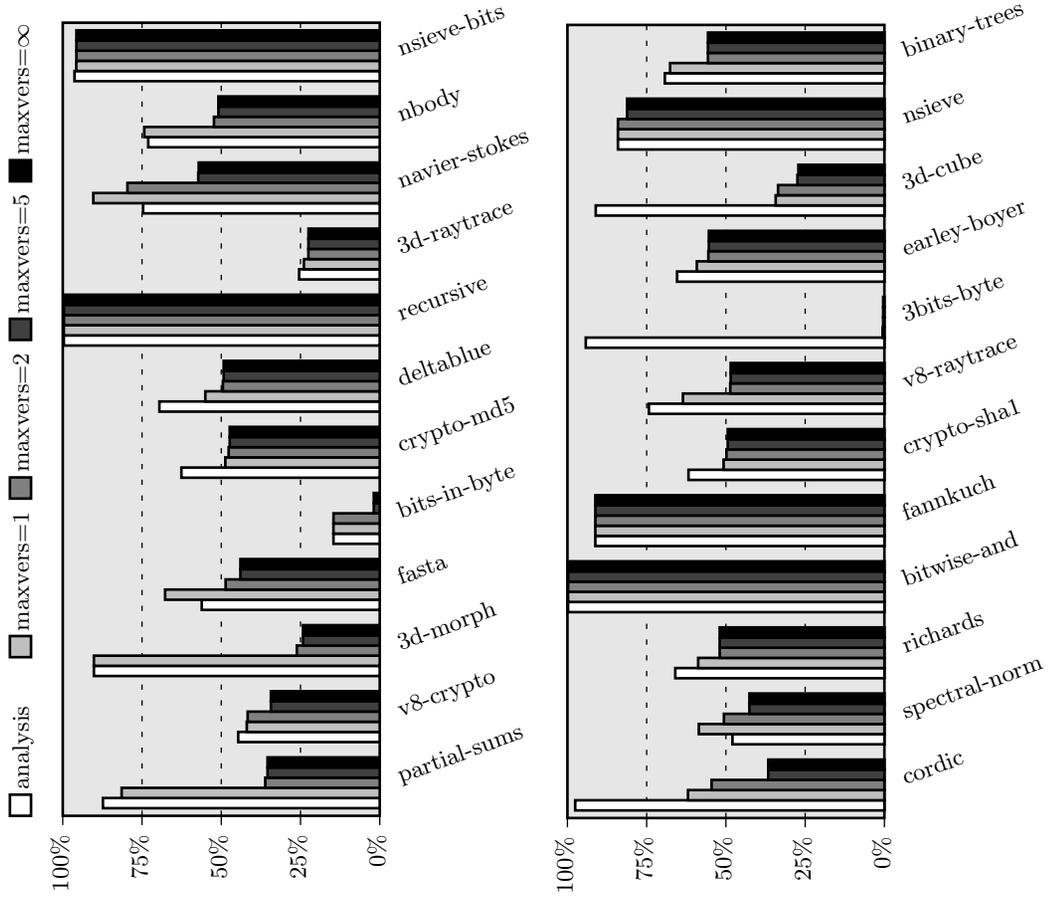

Figure \ref{fig:test_counts} shows counts of dynamically executed type tests
across all benchmarks for the representation analysis and for basic block
versioning with various block version limits. These counts are relative to
a baseline which has the version limit set to 0, and thus only generates a
default, unoptimized version of each basic block, without attempting to
eliminate any type tests. As can be seen from the counts, the
analysis produces a reduction in the number of dynamically executed type
tests over the unoptimized default on every benchmark. The basic block
versioning approach does at least as well as the analysis, and almost always
significantly better. Surprisingly, even with a version cap as low as 1
version per basic block, the versioning approach is often competitive with the
representation analysis.

\begin{figure}
\begin{centeredbox}
\begin{lstlisting}[language=javascript]
function bitsinbyte(b) {
    var m = 1, c = 0;
    while(m < 0x100) {
        if(b & m) c++;
        m <<= 1;
    }
    return c;
}

function TimeFunc(func) {
    var x, y, t;
    for(var x = 0; x < 350; x++)
    for(var y = 0; y < 256; y++) func(y);
}
TimeFunc(bitsinbyte);
\end{lstlisting}
\end{centeredbox}
\caption{SunSpider bits-in-byte benchmark\label{fig:bits-in-byte}}
\end{figure}

Raising the version cap reduces the number of tests performed with the
versioning approach in a seemingly asymptotic manner as we get closer to the
limit of what is achievable with our implementation. The versioning
approach does remarkably well on the {\tt bits-in-byte} benchmark, with a
reduction in the number of type tests by a factor of over 50. This benchmark
(see Figure \ref{fig:bits-in-byte}) is an ideal use case for our versioning
approach. It is a tight loop performing bitwise and arithmetic
operations on integers which are all stored in local variables. The versioning
approach performs noticeably better than the analysis on this test because it
is able to hoist a type test on the function parameter {\tt b} out of a critical
loop. The type of this parameter is initially unknown when entering the
function. The analysis on its own cannot achieve this, and so must repeat the
test every loop iteration. Note that neither the analysis nor the basic block
versioning approach need to test the type of {\tt c} at run time because the
variable is initialized to an integer value before loop entry, and integer overflow
never occurs, so the overflow case remains a stub. The {\tt bitwise-and}
benchmark operates exclusively on global variables, for which our system
cannot extract types, and so neither the type analysis nor the versioning
approach show any improvement over baseline for this benchmark.

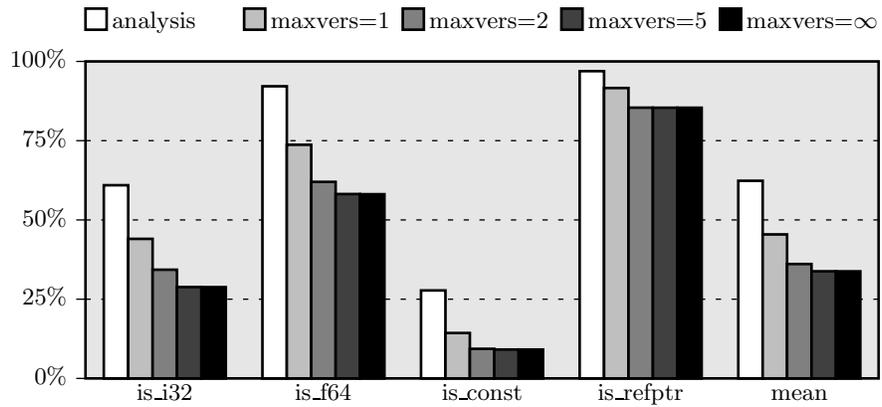
\begin{figure}
\begin{center}
\bgroup
\psset{unit=1pt}
\begin{pspicture}(329.,136.4)
\newgray{gray0}{0.}
\newgray{gray25}{.25}
\newgray{gray50}{.5}
\newgray{gray75}{.75}
\newgray{gray100}{1.}
\newgray{gray90}{.9}
\newgray{gray0}{0}
\psframe[linewidth=1.,linecolor=gray0,fillcolor=gray90,fillstyle=solid](27.5,-5.1000000000000005)(328.5,115.9)
\psframe[linewidth=1.,linecolor=gray0,fillcolor=gray100,fillstyle=solid](27.5,126.9)(36.5,135.9)
\rput[Bl]{0}(38.,128.2){analysis}
\psframe[linewidth=1.,linecolor=gray0,fillcolor=gray75,fillstyle=solid](87.5,126.9)(96.5,135.9)
\rput[Bl]{0}(98.,128.2){maxvers=1}
\psframe[linewidth=1.,linecolor=gray0,fillcolor=gray50,fillstyle=solid](147.5,126.9)(156.5,135.9)
\rput[Bl]{0}(158.,128.2){maxvers=2}
\psframe[linewidth=1.,linecolor=gray0,fillcolor=gray25,fillstyle=solid](207.5,126.9)(216.5,135.9)
\rput[Bl]{0}(218.,128.2){maxvers=5}
\psframe[linewidth=1.,linecolor=gray0,fillcolor=gray0,fillstyle=solid](267.5,126.9)(276.5,135.9)
\rput[Bl]{0}(278.,128.2){maxvers=$\infty$}
\psline[linewidth=.5,linecolor=gray0](24.5,-4.6000000000000005)(28.,-4.6000000000000005)
\rput[Br]{0}(21.,-6.7){0\%}
\psline[linewidth=.5,linecolor=gray0](24.5,25.400000000000002)(28.,25.400000000000002)
\psline[linewidth=.5,linecolor=gray0,linestyle=dashed,dash=2 4](28.,25.400000000000002)(328.,25.400000000000002)
\rput[Br]{0}(21.,23.3){25\%}
\psline[linewidth=.5,linecolor=gray0](24.5,55.4)(28.,55.4)
\psline[linewidth=.5,linecolor=gray0,linestyle=dashed,dash=2 4](28.,55.4)(328.,55.4)
\rput[Br]{0}(21.,53.300000000000004){50\%}
\psline[linewidth=.5,linecolor=gray0](24.5,85.4)(28.,85.4)
\psline[linewidth=.5,linecolor=gray0,linestyle=dashed,dash=2 4](28.,85.4)(328.,85.4)
\rput[Br]{0}(21.,83.3){75\%}
\psline[linewidth=.5,linecolor=gray0](24.5,115.4)(28.,115.4)
\rput[Br]{0}(21.,113.3){100\%}
\rput[B]{0}(58.,-12.6){is\_i32}
\psframe[linewidth=1.,linecolor=gray0,fillcolor=gray100,fillstyle=solid](34.423,-5.1000000000000005)(44.654,69.036)
\psframe[linewidth=1.,linecolor=gray0,fillcolor=gray75,fillstyle=solid](43.654,-5.1000000000000005)(53.885,48.728)
\psframe[linewidth=1.,linecolor=gray0,fillcolor=gray50,fillstyle=solid](52.885,-5.1000000000000005)(63.115,37.055)
\psframe[linewidth=1.,linecolor=gray0,fillcolor=gray25,fillstyle=solid](62.115,-5.1000000000000005)(72.346,30.490000000000002)
\psframe[linewidth=1.,linecolor=gray0,fillcolor=gray0,fillstyle=solid](71.346,-5.1000000000000005)(81.577,30.459)
\rput[B]{0}(118.,-12.6){is\_f64}
\psframe[linewidth=1.,linecolor=gray0,fillcolor=gray100,fillstyle=solid](94.423,-5.1000000000000005)(104.654,106.494)
\psframe[linewidth=1.,linecolor=gray0,fillcolor=gray75,fillstyle=solid](103.654,-5.1000000000000005)(113.885,84.351)
\psframe[linewidth=1.,linecolor=gray0,fillcolor=gray50,fillstyle=solid](112.885,-5.1000000000000005)(123.11500000000001,70.304)
\psframe[linewidth=1.,linecolor=gray0,fillcolor=gray25,fillstyle=solid](122.11500000000001,-5.1000000000000005)(132.346,65.699)
\psframe[linewidth=1.,linecolor=gray0,fillcolor=gray0,fillstyle=solid](131.346,-5.1000000000000005)(141.577,65.636)
\rput[B]{0}(178.,-12.6){is\_const}
\psframe[linewidth=1.,linecolor=gray0,fillcolor=gray100,fillstyle=solid](154.423,-5.1000000000000005)(164.654,29.206)
\psframe[linewidth=1.,linecolor=gray0,fillcolor=gray75,fillstyle=solid](163.654,-5.1000000000000005)(173.885,13.093)
\psframe[linewidth=1.,linecolor=gray0,fillcolor=gray50,fillstyle=solid](172.885,-5.1000000000000005)(183.115,7.109)
\psframe[linewidth=1.,linecolor=gray0,fillcolor=gray25,fillstyle=solid](182.115,-5.1000000000000005)(192.346,6.835)
\psframe[linewidth=1.,linecolor=gray0,fillcolor=gray0,fillstyle=solid](191.346,-5.1000000000000005)(201.577,6.835)
\rput[B]{0}(238.,-12.6){is\_refptr}
\psframe[linewidth=1.,linecolor=gray0,fillcolor=gray100,fillstyle=solid](214.423,-5.1000000000000005)(224.654,112.21900000000001)
\psframe[linewidth=1.,linecolor=gray0,fillcolor=gray75,fillstyle=solid](223.654,-5.1000000000000005)(233.885,105.839)
\psframe[linewidth=1.,linecolor=gray0,fillcolor=gray50,fillstyle=solid](232.885,-5.1000000000000005)(243.115,98.4)
\psframe[linewidth=1.,linecolor=gray0,fillcolor=gray25,fillstyle=solid](242.115,-5.1000000000000005)(252.346,98.374)
\psframe[linewidth=1.,linecolor=gray0,fillcolor=gray0,fillstyle=solid](251.346,-5.1000000000000005)(261.577,98.374)
\rput[B]{0}(298.,-12.6){mean}
\psframe[linewidth=1.,linecolor=gray0,fillcolor=gray100,fillstyle=solid](274.423,-5.1000000000000005)(284.654,70.71900000000001)
\psframe[linewidth=1.,linecolor=gray0,fillcolor=gray75,fillstyle=solid](283.654,-5.1000000000000005)(293.885,50.415)
\psframe[linewidth=1.,linecolor=gray0,fillcolor=gray50,fillstyle=solid](292.885,-5.1000000000000005)(303.115,39.209)
\psframe[linewidth=1.,linecolor=gray0,fillcolor=gray25,fillstyle=solid](302.115,-5.1000000000000005)(312.346,36.457)
\psframe[linewidth=1.,linecolor=gray0,fillcolor=gray0,fillstyle=solid](311.346,-5.1000000000000005)(321.577,36.438)
\end{pspicture}
\egroup
\end{center}
\caption{Type test counts by kind of type test (relative to baseline)\label{fig:test_kinds}}
\end{figure}

A breakdown of relative type test counts by kind, averaged accross all
benchmarks (using the geometric mean) is shown in Figure \ref{fig:test_kinds}.
We see that the versioning approach is able to achieve better results than the
representation analysis across each kind of type test. The {\tt is\_refptr}
category shows the least improvement. This is likely because property access
primitives are very large, and thus seldom inlined, limiting the ability of
both basic block versioning and the analysis to propagate type information for
reference values. We note that versioning is much more effective than the
analysis when it comes to eliminating {\tt is\_float64} type tests. This is
probably because integer and floating point types often get intermixed,
leading to cases where the analysis cannot eliminate such tests. The
versioning approach has the advantage that it can replicate and detangle
integer and floating point code paths. A limit of 5 versions per block
eliminates 64\unskip\% of type tests on average
(geometric mean), compared to 33\unskip\% for the
analysis.

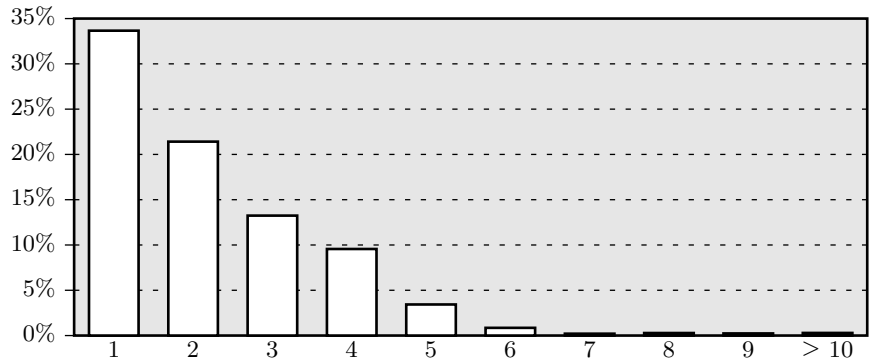
\begin{figure}
\begin{center}
\bgroup
\psset{unit=1pt}
\begin{pspicture}(324.8,116.4)
\newgray{gray100}{1.}
\newgray{gray90}{.9}
\newgray{gray0}{0}
\psframe[linewidth=1.,linecolor=gray0,fillcolor=gray90,fillstyle=solid](23.3,-5.1000000000000005)(324.3,115.9)
\psline[linewidth=.5,linecolor=gray0](20.3,-4.6000000000000005)(23.8,-4.6000000000000005)
\rput[Br]{0}(16.8,-6.7){0\%}
\psline[linewidth=.5,linecolor=gray0](20.3,12.543000000000001)(23.8,12.543000000000001)
\psline[linewidth=.5,linecolor=gray0,linestyle=dashed,dash=2 4](23.8,12.543000000000001)(323.8,12.543000000000001)
\rput[Br]{0}(16.8,10.443){5\%}
\psline[linewidth=.5,linecolor=gray0](20.3,29.686)(23.8,29.686)
\psline[linewidth=.5,linecolor=gray0,linestyle=dashed,dash=2 4](23.8,29.686)(323.8,29.686)
\rput[Br]{0}(16.8,27.586000000000002){10\%}
\psline[linewidth=.5,linecolor=gray0](20.3,46.829)(23.8,46.829)
\psline[linewidth=.5,linecolor=gray0,linestyle=dashed,dash=2 4](23.8,46.829)(323.8,46.829)
\rput[Br]{0}(16.8,44.729){15\%}
\psline[linewidth=.5,linecolor=gray0](20.3,63.971000000000004)(23.8,63.971000000000004)
\psline[linewidth=.5,linecolor=gray0,linestyle=dashed,dash=2 4](23.8,63.971000000000004)(323.8,63.971000000000004)
\rput[Br]{0}(16.8,61.871){20\%}
\psline[linewidth=.5,linecolor=gray0](20.3,81.114)(23.8,81.114)
\psline[linewidth=.5,linecolor=gray0,linestyle=dashed,dash=2 4](23.8,81.114)(323.8,81.114)
\rput[Br]{0}(16.8,79.014){25\%}
\psline[linewidth=.5,linecolor=gray0](20.3,98.257)(23.8,98.257)
\psline[linewidth=.5,linecolor=gray0,linestyle=dashed,dash=2 4](23.8,98.257)(323.8,98.257)
\rput[Br]{0}(16.8,96.157){30\%}
\psline[linewidth=.5,linecolor=gray0](20.3,115.4)(23.8,115.4)
\rput[Br]{0}(16.8,113.3){35\%}
\rput[B]{0}(38.800000000000004,-12.6){1}
\psframe[linewidth=1.,linecolor=gray0,fillcolor=gray100,fillstyle=solid](28.925,-5.1000000000000005)(48.675000000000004,111.316)
\rput[B]{0}(68.8,-12.6){2}
\psframe[linewidth=1.,linecolor=gray0,fillcolor=gray100,fillstyle=solid](58.925000000000004,-5.1000000000000005)(78.675,69.312)
\rput[B]{0}(98.8,-12.6){3}
\psframe[linewidth=1.,linecolor=gray0,fillcolor=gray100,fillstyle=solid](88.925,-5.1000000000000005)(108.675,41.306)
\rput[B]{0}(128.8,-12.6){4}
\psframe[linewidth=1.,linecolor=gray0,fillcolor=gray100,fillstyle=solid](118.925,-5.1000000000000005)(138.675,28.67)
\rput[B]{0}(158.8,-12.6){5}
\psframe[linewidth=1.,linecolor=gray0,fillcolor=gray100,fillstyle=solid](148.925,-5.1000000000000005)(168.675,7.666)
\rput[B]{0}(188.8,-12.6){6}
\psframe[linewidth=1.,linecolor=gray0,fillcolor=gray100,fillstyle=solid](178.925,-5.1000000000000005)(198.675,-1.179)
\rput[B]{0}(218.8,-12.6){7}
\psframe[linewidth=1.,linecolor=gray0,fillcolor=gray100,fillstyle=solid](208.925,-5.1000000000000005)(228.675,-3.392)
\rput[B]{0}(248.8,-12.6){8}
\psframe[linewidth=1.,linecolor=gray0,fillcolor=gray100,fillstyle=solid](238.925,-5.1000000000000005)(258.675,-3.1390000000000002)
\rput[B]{0}(278.8,-12.6){9}
\psframe[linewidth=1.,linecolor=gray0,fillcolor=gray100,fillstyle=solid](268.925,-5.1000000000000005)(288.675,-3.2920000000000003)
\rput[B]{0}(308.8,-12.6){$\ge$ 10}
\psframe[linewidth=1.,linecolor=gray0,fillcolor=gray100,fillstyle=solid](298.925,-5.1000000000000005)(318.675,-3.108)
\end{pspicture}
\egroup
\end{center}
\caption{Relative occurrence of block version counts\label{fig:ver_counts}}
\end{figure}

Figure \ref{fig:ver_counts} shows the relative proportion of blocks for which
different numbers of versions were generated, averaged accross all benchmarks
(geometric mean). As one might expect, the relative proportion of blocks tends
to steadily decrease as the number of versions is increased. Most blocks only have one
or two versions, and less than 9\% have 5 versions or more. There are very few
blocks which have 10 versions or more. These are a small minority, but such
pathological cases do occur in practice.

The function generating the most block versions in our tests is {\tt DrawLine}
from the {\tt 3d-cube} benchmark, which produces 32 versions of
one particular block. This function draws a line in screen space between point
coordinates {\tt x1,y1} and {\tt x2,y2}. Multiple different values are computed
inside {\tt DrawLine} based on these points. Each of the coordinate values can
be either integer or  floating point, which results in a situation where there
are several live variables, all of which can have two different types. This
creates an explosion in the number of versions of blocks inside this function
as basic block versioning tries to account for all possible type combinations
of these values. In practice, the values are either all integer, or
all floating point, but our implementation of basic block versioning is
currently unable to take advantage of this helpful fact. We have experimentally
verified that, in fact, only 17 of the 32 versions generated in {\tt DrawLine}
are actually executed. A strategy for addressing this problem is discussed in
Section \ref{sec:future}.

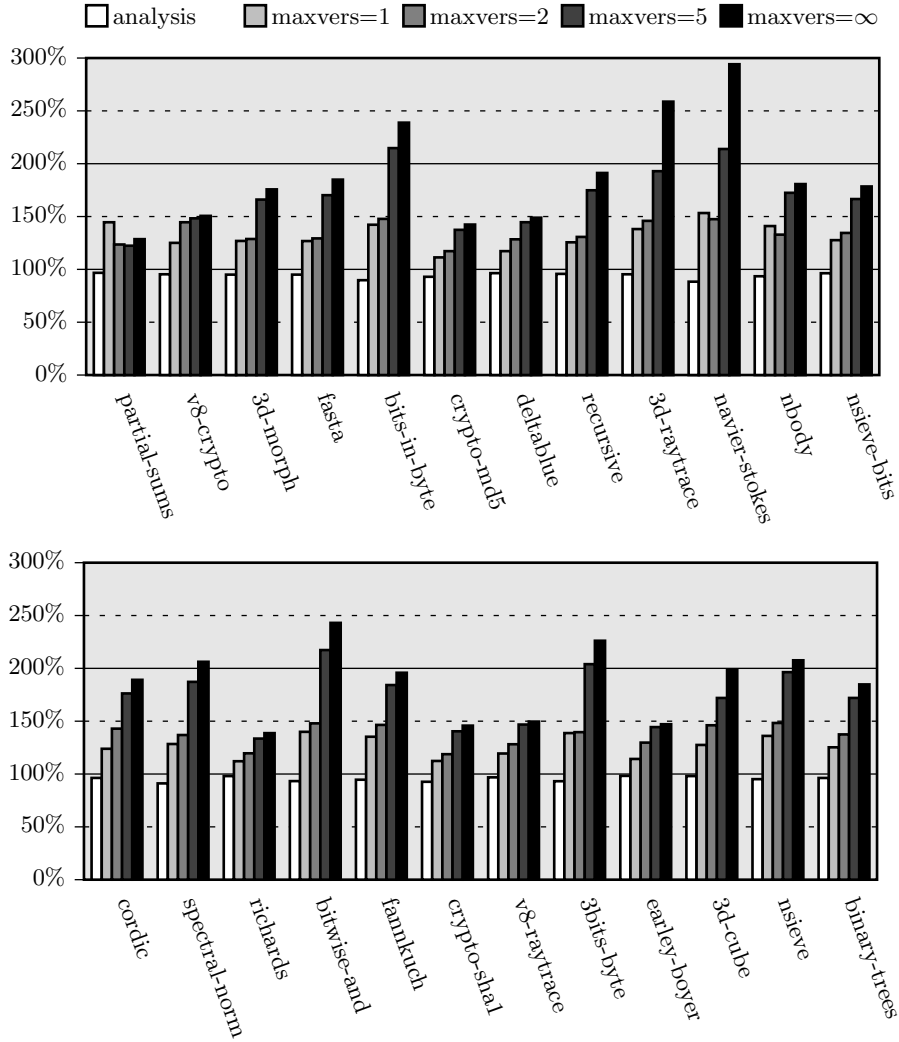
\begin{figure}
\begin{center}
\bgroup
\psset{unit=1pt}
\begin{pspicture}(334.559,210.037)
\newgray{gray0}{0.}
\newgray{gray25}{.25}
\newgray{gray50}{.5}
\newgray{gray75}{.75}
\newgray{gray100}{1.}
\newgray{gray90}{.9}
\newgray{gray0}{0}
\psframe[linewidth=1.,linecolor=gray0,fillcolor=gray90,fillstyle=solid](27.5,68.537)(328.5,189.537)
\psframe[linewidth=1.,linecolor=gray0,fillcolor=gray100,fillstyle=solid](27.5,200.537)(36.5,209.537)
\rput[Bl]{0}(38.,201.83700000000002){analysis}
\psframe[linewidth=1.,linecolor=gray0,fillcolor=gray75,fillstyle=solid](87.5,200.537)(96.5,209.537)
\rput[Bl]{0}(98.,201.83700000000002){maxvers=1}
\psframe[linewidth=1.,linecolor=gray0,fillcolor=gray50,fillstyle=solid](147.5,200.537)(156.5,209.537)
\rput[Bl]{0}(158.,201.83700000000002){maxvers=2}
\psframe[linewidth=1.,linecolor=gray0,fillcolor=gray25,fillstyle=solid](207.5,200.537)(216.5,209.537)
\rput[Bl]{0}(218.,201.83700000000002){maxvers=5}
\psframe[linewidth=1.,linecolor=gray0,fillcolor=gray0,fillstyle=solid](267.5,200.537)(276.5,209.537)
\rput[Bl]{0}(278.,201.83700000000002){maxvers=$\infty$}
\psline[linewidth=.5,linecolor=gray0](24.5,69.037)(28.,69.037)
\rput[Br]{0}(21.,66.937){0\%}
\psline[linewidth=.5,linecolor=gray0](24.5,89.037)(28.,89.037)
\psline[linewidth=.5,linecolor=gray0,linestyle=dashed,dash=2 4](28.,89.037)(328.,89.037)
\rput[Br]{0}(21.,86.937){50\%}
\psline[linewidth=.5,linecolor=gray0](24.5,109.037)(28.,109.037)
\psline[linewidth=.5,linecolor=gray0](28.,109.037)(328.,109.037)
\rput[Br]{0}(21.,106.937){100\%}
\psline[linewidth=.5,linecolor=gray0](24.5,129.037)(28.,129.037)
\psline[linewidth=.5,linecolor=gray0,linestyle=dashed,dash=2 4](28.,129.037)(328.,129.037)
\rput[Br]{0}(21.,126.937){150\%}
\psline[linewidth=.5,linecolor=gray0](24.5,149.037)(28.,149.037)
\psline[linewidth=.5,linecolor=gray0](28.,149.037)(328.,149.037)
\rput[Br]{0}(21.,146.937){200\%}
\psline[linewidth=.5,linecolor=gray0](24.5,169.037)(28.,169.037)
\psline[linewidth=.5,linecolor=gray0,linestyle=dashed,dash=2 4](28.,169.037)(328.,169.037)
\rput[Br]{0}(21.,166.937){250\%}
\psline[linewidth=.5,linecolor=gray0](24.5,189.037)(28.,189.037)
\rput[Br]{0}(21.,186.937){300\%}
\rput[Bl]{-70.}(40.5,61.037){partial-sums}
\psframe[linewidth=1.,linecolor=gray0,fillcolor=gray100,fillstyle=solid](30.385,68.537)(35.231,108.226)
\psframe[linewidth=1.,linecolor=gray0,fillcolor=gray75,fillstyle=solid](34.231,68.537)(39.077,127.37100000000001)
\psframe[linewidth=1.,linecolor=gray0,fillcolor=gray50,fillstyle=solid](38.077,68.537)(42.923,118.949)
\psframe[linewidth=1.,linecolor=gray0,fillcolor=gray25,fillstyle=solid](41.923,68.537)(46.769,118.494)
\psframe[linewidth=1.,linecolor=gray0,fillcolor=gray0,fillstyle=solid](45.769,68.537)(50.615,120.979)
\rput[Bl]{-70.}(65.5,61.037){v8-crypto}
\psframe[linewidth=1.,linecolor=gray0,fillcolor=gray100,fillstyle=solid](55.385,68.537)(60.231,107.676)
\psframe[linewidth=1.,linecolor=gray0,fillcolor=gray75,fillstyle=solid](59.231,68.537)(64.077,119.572)
\psframe[linewidth=1.,linecolor=gray0,fillcolor=gray50,fillstyle=solid](63.077,68.537)(67.923,127.373)
\psframe[linewidth=1.,linecolor=gray0,fillcolor=gray25,fillstyle=solid](66.923,68.537)(71.769,128.853)
\psframe[linewidth=1.,linecolor=gray0,fillcolor=gray0,fillstyle=solid](70.769,68.537)(75.615,129.792)
\rput[Bl]{-70.}(90.5,61.037){3d-morph}
\psframe[linewidth=1.,linecolor=gray0,fillcolor=gray100,fillstyle=solid](80.385,68.537)(85.23100000000001,107.549)
\psframe[linewidth=1.,linecolor=gray0,fillcolor=gray75,fillstyle=solid](84.23100000000001,68.537)(89.077,120.297)
\psframe[linewidth=1.,linecolor=gray0,fillcolor=gray50,fillstyle=solid](88.077,68.537)(92.923,121.035)
\psframe[linewidth=1.,linecolor=gray0,fillcolor=gray25,fillstyle=solid](91.923,68.537)(96.769,135.961)
\psframe[linewidth=1.,linecolor=gray0,fillcolor=gray0,fillstyle=solid](95.769,68.537)(100.61500000000001,139.818)
\rput[Bl]{-70.}(115.5,61.037){fasta}
\psframe[linewidth=1.,linecolor=gray0,fillcolor=gray100,fillstyle=solid](105.385,68.537)(110.23100000000001,107.549)
\psframe[linewidth=1.,linecolor=gray0,fillcolor=gray75,fillstyle=solid](109.23100000000001,68.537)(114.077,120.256)
\psframe[linewidth=1.,linecolor=gray0,fillcolor=gray50,fillstyle=solid](113.077,68.537)(117.923,121.26)
\psframe[linewidth=1.,linecolor=gray0,fillcolor=gray25,fillstyle=solid](116.923,68.537)(121.769,137.62)
\psframe[linewidth=1.,linecolor=gray0,fillcolor=gray0,fillstyle=solid](120.769,68.537)(125.61500000000001,143.51)
\rput[Bl]{-70.}(140.5,61.037){bits-in-byte}
\psframe[linewidth=1.,linecolor=gray0,fillcolor=gray100,fillstyle=solid](130.385,68.537)(135.231,105.441)
\psframe[linewidth=1.,linecolor=gray0,fillcolor=gray75,fillstyle=solid](134.231,68.537)(139.077,126.434)
\psframe[linewidth=1.,linecolor=gray0,fillcolor=gray50,fillstyle=solid](138.077,68.537)(142.923,128.647)
\psframe[linewidth=1.,linecolor=gray0,fillcolor=gray25,fillstyle=solid](141.923,68.537)(146.769,155.443)
\psframe[linewidth=1.,linecolor=gray0,fillcolor=gray0,fillstyle=solid](145.769,68.537)(150.615,165.079)
\rput[Bl]{-70.}(165.5,61.037){crypto-md5}
\psframe[linewidth=1.,linecolor=gray0,fillcolor=gray100,fillstyle=solid](155.385,68.537)(160.231,106.72800000000001)
\psframe[linewidth=1.,linecolor=gray0,fillcolor=gray75,fillstyle=solid](159.231,68.537)(164.077,114.066)
\psframe[linewidth=1.,linecolor=gray0,fillcolor=gray50,fillstyle=solid](163.077,68.537)(167.923,116.433)
\psframe[linewidth=1.,linecolor=gray0,fillcolor=gray25,fillstyle=solid](166.923,68.537)(171.769,124.501)
\psframe[linewidth=1.,linecolor=gray0,fillcolor=gray0,fillstyle=solid](170.769,68.537)(175.615,126.501)
\rput[Bl]{-70.}(190.5,61.037){deltablue}
\psframe[linewidth=1.,linecolor=gray0,fillcolor=gray100,fillstyle=solid](180.385,68.537)(185.231,108.102)
\psframe[linewidth=1.,linecolor=gray0,fillcolor=gray75,fillstyle=solid](184.231,68.537)(189.077,116.437)
\psframe[linewidth=1.,linecolor=gray0,fillcolor=gray50,fillstyle=solid](188.077,68.537)(192.923,120.913)
\psframe[linewidth=1.,linecolor=gray0,fillcolor=gray25,fillstyle=solid](191.923,68.537)(196.769,127.382)
\psframe[linewidth=1.,linecolor=gray0,fillcolor=gray0,fillstyle=solid](195.769,68.537)(200.615,129.095)
\rput[Bl]{-70.}(215.5,61.037){recursive}
\psframe[linewidth=1.,linecolor=gray0,fillcolor=gray100,fillstyle=solid](205.385,68.537)(210.231,107.831)
\psframe[linewidth=1.,linecolor=gray0,fillcolor=gray75,fillstyle=solid](209.231,68.537)(214.077,119.782)
\psframe[linewidth=1.,linecolor=gray0,fillcolor=gray50,fillstyle=solid](213.077,68.537)(217.923,121.82300000000001)
\psframe[linewidth=1.,linecolor=gray0,fillcolor=gray25,fillstyle=solid](216.923,68.537)(221.769,139.482)
\psframe[linewidth=1.,linecolor=gray0,fillcolor=gray0,fillstyle=solid](220.769,68.537)(225.615,146.03)
\rput[Bl]{-70.}(240.5,61.037){3d-raytrace}
\psframe[linewidth=1.,linecolor=gray0,fillcolor=gray100,fillstyle=solid](230.385,68.537)(235.231,107.675)
\psframe[linewidth=1.,linecolor=gray0,fillcolor=gray75,fillstyle=solid](234.231,68.537)(239.077,124.794)
\psframe[linewidth=1.,linecolor=gray0,fillcolor=gray50,fillstyle=solid](238.077,68.537)(242.923,127.903)
\psframe[linewidth=1.,linecolor=gray0,fillcolor=gray25,fillstyle=solid](241.923,68.537)(246.769,146.704)
\psframe[linewidth=1.,linecolor=gray0,fillcolor=gray0,fillstyle=solid](245.769,68.537)(250.615,173.032)
\rput[Bl]{-70.}(265.5,61.037){navier-stokes}
\psframe[linewidth=1.,linecolor=gray0,fillcolor=gray100,fillstyle=solid](255.38500000000002,68.537)(260.231,104.876)
\psframe[linewidth=1.,linecolor=gray0,fillcolor=gray75,fillstyle=solid](259.231,68.537)(264.077,130.865)
\psframe[linewidth=1.,linecolor=gray0,fillcolor=gray50,fillstyle=solid](263.077,68.537)(267.923,128.54)
\psframe[linewidth=1.,linecolor=gray0,fillcolor=gray25,fillstyle=solid](266.923,68.537)(271.769,155.1)
\psframe[linewidth=1.,linecolor=gray0,fillcolor=gray0,fillstyle=solid](270.769,68.537)(275.615,187.193)
\rput[Bl]{-70.}(290.5,61.037){nbody}
\psframe[linewidth=1.,linecolor=gray0,fillcolor=gray100,fillstyle=solid](280.385,68.537)(285.231,106.927)
\psframe[linewidth=1.,linecolor=gray0,fillcolor=gray75,fillstyle=solid](284.231,68.537)(289.077,125.935)
\psframe[linewidth=1.,linecolor=gray0,fillcolor=gray50,fillstyle=solid](288.077,68.537)(292.923,122.684)
\psframe[linewidth=1.,linecolor=gray0,fillcolor=gray25,fillstyle=solid](291.923,68.537)(296.769,138.52)
\psframe[linewidth=1.,linecolor=gray0,fillcolor=gray0,fillstyle=solid](295.769,68.537)(300.615,141.828)
\rput[Bl]{-70.}(315.5,61.037){nsieve-bits}
\psframe[linewidth=1.,linecolor=gray0,fillcolor=gray100,fillstyle=solid](305.385,68.537)(310.231,108.07600000000001)
\psframe[linewidth=1.,linecolor=gray0,fillcolor=gray75,fillstyle=solid](309.231,68.537)(314.077,120.598)
\psframe[linewidth=1.,linecolor=gray0,fillcolor=gray50,fillstyle=solid](313.077,68.537)(317.923,123.33)
\psframe[linewidth=1.,linecolor=gray0,fillcolor=gray25,fillstyle=solid](316.923,68.537)(321.769,136.159)
\psframe[linewidth=1.,linecolor=gray0,fillcolor=gray0,fillstyle=solid](320.769,68.537)(325.615,140.915)
\end{pspicture}
\egroup
\bgroup
\psset{unit=1pt}
\begin{pspicture}(336.2,170.037)
\newgray{gray0}{0.}
\newgray{gray25}{.25}
\newgray{gray50}{.5}
\newgray{gray75}{.75}
\newgray{gray100}{1.}
\newgray{gray90}{.9}
\newgray{gray0}{0}
\psframe[linewidth=1.,linecolor=gray0,fillcolor=gray90,fillstyle=solid](27.5,48.537)(328.5,169.537)
\psline[linewidth=.5,linecolor=gray0](24.5,49.037)(28.,49.037)
\rput[Br]{0}(21.,46.937){0\%}
\psline[linewidth=.5,linecolor=gray0](24.5,69.037)(28.,69.037)
\psline[linewidth=.5,linecolor=gray0,linestyle=dashed,dash=2 4](28.,69.037)(328.,69.037)
\rput[Br]{0}(21.,66.937){50\%}
\psline[linewidth=.5,linecolor=gray0](24.5,89.037)(28.,89.037)
\psline[linewidth=.5,linecolor=gray0](28.,89.037)(328.,89.037)
\rput[Br]{0}(21.,86.937){100\%}
\psline[linewidth=.5,linecolor=gray0](24.5,109.037)(28.,109.037)
\psline[linewidth=.5,linecolor=gray0,linestyle=dashed,dash=2 4](28.,109.037)(328.,109.037)
\rput[Br]{0}(21.,106.937){150\%}
\psline[linewidth=.5,linecolor=gray0](24.5,129.037)(28.,129.037)
\psline[linewidth=.5,linecolor=gray0](28.,129.037)(328.,129.037)
\rput[Br]{0}(21.,126.937){200\%}
\psline[linewidth=.5,linecolor=gray0](24.5,149.037)(28.,149.037)
\psline[linewidth=.5,linecolor=gray0,linestyle=dashed,dash=2 4](28.,149.037)(328.,149.037)
\rput[Br]{0}(21.,146.937){250\%}
\psline[linewidth=.5,linecolor=gray0](24.5,169.037)(28.,169.037)
\rput[Br]{0}(21.,166.937){300\%}
\rput[Bl]{-70.}(40.5,41.037){cordic}
\psframe[linewidth=1.,linecolor=gray0,fillcolor=gray100,fillstyle=solid](30.385,48.537)(35.231,88.059)
\psframe[linewidth=1.,linecolor=gray0,fillcolor=gray75,fillstyle=solid](34.231,48.537)(39.077,99.096)
\psframe[linewidth=1.,linecolor=gray0,fillcolor=gray50,fillstyle=solid](38.077,48.537)(42.923,106.718)
\psframe[linewidth=1.,linecolor=gray0,fillcolor=gray25,fillstyle=solid](41.923,48.537)(46.769,120.041)
\psframe[linewidth=1.,linecolor=gray0,fillcolor=gray0,fillstyle=solid](45.769,48.537)(50.615,125.196)
\rput[Bl]{-70.}(65.5,41.037){spectral-norm}
\psframe[linewidth=1.,linecolor=gray0,fillcolor=gray100,fillstyle=solid](55.385,48.537)(60.231,85.989)
\psframe[linewidth=1.,linecolor=gray0,fillcolor=gray75,fillstyle=solid](59.231,48.537)(64.077,100.9)
\psframe[linewidth=1.,linecolor=gray0,fillcolor=gray50,fillstyle=solid](63.077,48.537)(67.923,104.303)
\psframe[linewidth=1.,linecolor=gray0,fillcolor=gray25,fillstyle=solid](66.923,48.537)(71.769,124.424)
\psframe[linewidth=1.,linecolor=gray0,fillcolor=gray0,fillstyle=solid](70.769,48.537)(75.615,132.024)
\rput[Bl]{-70.}(90.5,41.037){richards}
\psframe[linewidth=1.,linecolor=gray0,fillcolor=gray100,fillstyle=solid](80.385,48.537)(85.23100000000001,88.757)
\psframe[linewidth=1.,linecolor=gray0,fillcolor=gray75,fillstyle=solid](84.23100000000001,48.537)(89.077,94.383)
\psframe[linewidth=1.,linecolor=gray0,fillcolor=gray50,fillstyle=solid](88.077,48.537)(92.923,97.38)
\psframe[linewidth=1.,linecolor=gray0,fillcolor=gray25,fillstyle=solid](91.923,48.537)(96.769,102.95)
\psframe[linewidth=1.,linecolor=gray0,fillcolor=gray0,fillstyle=solid](95.769,48.537)(100.61500000000001,105.015)
\rput[Bl]{-70.}(115.5,41.037){bitwise-and}
\psframe[linewidth=1.,linecolor=gray0,fillcolor=gray100,fillstyle=solid](105.385,48.537)(110.23100000000001,86.857)
\psframe[linewidth=1.,linecolor=gray0,fillcolor=gray75,fillstyle=solid](109.23100000000001,48.537)(114.077,105.51)
\psframe[linewidth=1.,linecolor=gray0,fillcolor=gray50,fillstyle=solid](113.077,48.537)(117.923,108.71600000000001)
\psframe[linewidth=1.,linecolor=gray0,fillcolor=gray25,fillstyle=solid](116.923,48.537)(121.769,136.466)
\psframe[linewidth=1.,linecolor=gray0,fillcolor=gray0,fillstyle=solid](120.769,48.537)(125.61500000000001,146.755)
\rput[Bl]{-70.}(140.5,41.037){fannkuch}
\psframe[linewidth=1.,linecolor=gray0,fillcolor=gray100,fillstyle=solid](130.385,48.537)(135.231,87.382)
\psframe[linewidth=1.,linecolor=gray0,fillcolor=gray75,fillstyle=solid](134.231,48.537)(139.077,103.648)
\psframe[linewidth=1.,linecolor=gray0,fillcolor=gray50,fillstyle=solid](138.077,48.537)(142.923,108.129)
\psframe[linewidth=1.,linecolor=gray0,fillcolor=gray25,fillstyle=solid](141.923,48.537)(146.769,123.21900000000001)
\psframe[linewidth=1.,linecolor=gray0,fillcolor=gray0,fillstyle=solid](145.769,48.537)(150.615,127.875)
\rput[Bl]{-70.}(165.5,41.037){crypto-sha1}
\psframe[linewidth=1.,linecolor=gray0,fillcolor=gray100,fillstyle=solid](155.385,48.537)(160.231,86.60000000000001)
\psframe[linewidth=1.,linecolor=gray0,fillcolor=gray75,fillstyle=solid](159.231,48.537)(164.077,94.48100000000001)
\psframe[linewidth=1.,linecolor=gray0,fillcolor=gray50,fillstyle=solid](163.077,48.537)(167.923,97.044)
\psframe[linewidth=1.,linecolor=gray0,fillcolor=gray25,fillstyle=solid](166.923,48.537)(171.769,105.70700000000001)
\psframe[linewidth=1.,linecolor=gray0,fillcolor=gray0,fillstyle=solid](170.769,48.537)(175.615,107.854)
\rput[Bl]{-70.}(190.5,41.037){v8-raytrace}
\psframe[linewidth=1.,linecolor=gray0,fillcolor=gray100,fillstyle=solid](180.385,48.537)(185.231,88.277)
\psframe[linewidth=1.,linecolor=gray0,fillcolor=gray75,fillstyle=solid](184.231,48.537)(189.077,97.304)
\psframe[linewidth=1.,linecolor=gray0,fillcolor=gray50,fillstyle=solid](188.077,48.537)(192.923,100.807)
\psframe[linewidth=1.,linecolor=gray0,fillcolor=gray25,fillstyle=solid](191.923,48.537)(196.769,108.26)
\psframe[linewidth=1.,linecolor=gray0,fillcolor=gray0,fillstyle=solid](195.769,48.537)(200.615,109.381)
\rput[Bl]{-70.}(215.5,41.037){3bits-byte}
\psframe[linewidth=1.,linecolor=gray0,fillcolor=gray100,fillstyle=solid](205.385,48.537)(210.231,86.813)
\psframe[linewidth=1.,linecolor=gray0,fillcolor=gray75,fillstyle=solid](209.231,48.537)(214.077,105.037)
\psframe[linewidth=1.,linecolor=gray0,fillcolor=gray50,fillstyle=solid](213.077,48.537)(217.923,105.386)
\psframe[linewidth=1.,linecolor=gray0,fillcolor=gray25,fillstyle=solid](216.923,48.537)(221.769,131.125)
\psframe[linewidth=1.,linecolor=gray0,fillcolor=gray0,fillstyle=solid](220.769,48.537)(225.615,139.992)
\rput[Bl]{-70.}(240.5,41.037){earley-boyer}
\psframe[linewidth=1.,linecolor=gray0,fillcolor=gray100,fillstyle=solid](230.385,48.537)(235.231,88.837)
\psframe[linewidth=1.,linecolor=gray0,fillcolor=gray75,fillstyle=solid](234.231,48.537)(239.077,95.241)
\psframe[linewidth=1.,linecolor=gray0,fillcolor=gray50,fillstyle=solid](238.077,48.537)(242.923,101.425)
\psframe[linewidth=1.,linecolor=gray0,fillcolor=gray25,fillstyle=solid](241.923,48.537)(246.769,107.298)
\psframe[linewidth=1.,linecolor=gray0,fillcolor=gray0,fillstyle=solid](245.769,48.537)(250.615,108.383)
\rput[Bl]{-70.}(265.5,41.037){3d-cube}
\psframe[linewidth=1.,linecolor=gray0,fillcolor=gray100,fillstyle=solid](255.38500000000002,48.537)(260.231,88.758)
\psframe[linewidth=1.,linecolor=gray0,fillcolor=gray75,fillstyle=solid](259.231,48.537)(264.077,100.544)
\psframe[linewidth=1.,linecolor=gray0,fillcolor=gray50,fillstyle=solid](263.077,48.537)(267.923,107.99300000000001)
\psframe[linewidth=1.,linecolor=gray0,fillcolor=gray25,fillstyle=solid](266.923,48.537)(271.769,118.33)
\psframe[linewidth=1.,linecolor=gray0,fillcolor=gray0,fillstyle=solid](270.769,48.537)(275.615,128.958)
\rput[Bl]{-70.}(290.5,41.037){nsieve}
\psframe[linewidth=1.,linecolor=gray0,fillcolor=gray100,fillstyle=solid](280.385,48.537)(285.231,87.59400000000001)
\psframe[linewidth=1.,linecolor=gray0,fillcolor=gray75,fillstyle=solid](284.231,48.537)(289.077,103.968)
\psframe[linewidth=1.,linecolor=gray0,fillcolor=gray50,fillstyle=solid](288.077,48.537)(292.923,108.861)
\psframe[linewidth=1.,linecolor=gray0,fillcolor=gray25,fillstyle=solid](291.923,48.537)(296.769,128.096)
\psframe[linewidth=1.,linecolor=gray0,fillcolor=gray0,fillstyle=solid](295.769,48.537)(300.615,132.586)
\rput[Bl]{-70.}(315.5,41.037){binary-trees}
\psframe[linewidth=1.,linecolor=gray0,fillcolor=gray100,fillstyle=solid](305.385,48.537)(310.231,88.044)
\psframe[linewidth=1.,linecolor=gray0,fillcolor=gray75,fillstyle=solid](309.231,48.537)(314.077,99.648)
\psframe[linewidth=1.,linecolor=gray0,fillcolor=gray50,fillstyle=solid](313.077,48.537)(317.923,104.53)
\psframe[linewidth=1.,linecolor=gray0,fillcolor=gray25,fillstyle=solid](316.923,48.537)(321.769,118.337)
\psframe[linewidth=1.,linecolor=gray0,fillcolor=gray0,fillstyle=solid](320.769,48.537)(325.615,123.456)
\end{pspicture}
\egroup
\end{center}
\caption{Code size growth for different block version limits\label{fig:code_size}}
\end{figure}

The effects of basic block versioning on the total generated code size are
shown in Figure \ref{fig:code_size}. It is interesting to note that the
representation analysis almost always results in a slight reduction in code size.
This is because the analysis allows the elimination of type tests and the
generation of more optimized code, which is often smaller. On the other hand,
basic block versioning can generate multiple versions of basic blocks,
which results in more generated code. The volume of generated code does not
increase linearly with the block version cap. Rather, it tapers off as a
limited number of versions tends to be generated for each block. Even without a
block version limit, the code size is less than double that of the baseline in
most cases. A limit of 5 versions per block results in a mean code size
increase of 69\unskip\%.

\section{Related Work}\label{sec:related}
There have been multiple efforts to devise type analyses for dynamic languages.
The Rapid Atomic Type Analysis (RATA)~\cite{rata} is an intraprocedural
flow-sensitive analysis based on abstract interpretation which aims to assign
unique types to each variable inside of a function. Attempts have
also been made to define formal semantics for a subset of dynamic languages
such as JavaScript~\cite{ti_js}, Ruby~\cite{ti_ruby} and Python~\cite{rpython},
sidestepping some of the complexity of these languages and making them more
amenable to traditional type inference techniques. There are
also flow-based interprocedural type analyses for JavaScript based on
sophisticated type lattices~\cite{tajs}\cite{tajs_lazy}. Such analyses are
usable in the context of static code analysis, but take too long to execute
to be usable in compilation and do not deal with the complexities of
dynamic code loading.

More recently, work done by Brian Hackett et al. resulted in an
interprocedural hybrid type analysis for JavaScript suitable for use in
production JIT compilers~\cite{mozti}. This analysis represents a great step
forward for dynamic languages, but as with other type analyses, must
assign one type to each value, which makes it vulnerable to imprecise type
information polluting analysis results. Basic block versioning could
potentially help improve on the results of such an analysis by hoisting tests
out of loops and generating multiple optimized code paths where appropriate.

Trace compilation aims to record long sequences of instructions executed inside
of hot loops~\cite{hotpathvm}. Such sequences of instructions often make
optimization easier. Type information can be accumulated along traces and used
to specialize the code to remove type tests~\cite{trace_spec}, overflow
checks~\cite{trace_ovf} and unnecessary allocations~\cite{trace_alloc}. Tracing
is similar to basic block versioning, in that context updating works on essentially linear
code fragments and accumulates type information during compilation. However, 
trace compilation incurs several difficulties and corner cases in practice,
such as the potential for trace explosion if there is a large number of
control-flow paths going through a loop, and poor capability to deal with code
that is not loop-based. Work on trace regions by Bebenita et al.~\cite{bebenita}
introduces traces with join nodes. These join nodes can potentially eliminate
tail duplication among traces and avoid the problem of trace explosion, but
also makes the compiler architecture more complex.

Basic block versioning bears some similarities to classic compiler
optimizations such as loop unrolling~\cite{loop_unrolling}, loop
peeling~\cite{loop_peeling}, and tail duplication, in that it achieves some of
the same results. Tail duplication and loop peeling are used in the formation
of hyperblocks~\cite{hyperblock}, which are sets of basic blocks grouped
together, such that control-flow may enter only into one of the blocks, but
may exit at multiple locations. This structure was designed to facilitate
the optimization of large units of code for VLIW architectures.
A parallel can be drawn between basic block versioning and Partial Redundancy
Elimination (PRE)~\cite{pre} in that the versioning approach seeks to eliminate
and hoists out of loops a specific kind of redundant computation, that of dynamic
type tests.

Basic block versioning is also similar to the idea of node
splitting~\cite{node_splitting}. This technique is an analysis device designed
to make control-flow graphs reducible and more amenable to analysis.
The path splitting algorithm implemented in the SUIF compiler~\cite{path_splitting}
aims at improving reaching definition information by replicating control-flow
nodes in loops to eliminate joins. Unlike basic block versioning, these
algorithm cannot gain information from type tests. The algorithms as presented
are also specifically targeted at loops, while basic block versioning makes no
special distinction. Similarly, a static analysis which replicates code to
eliminate conditional branches has been developed~\cite{code_replication}. 
This algorithm operates on a low-level intermediate representation, is
intended to optimize loops and does not specifically eliminate type tests.

Customization is a technique developed to optimize the SELF programming
language~\cite{self_customization} which compiles multiple copies of methods,
specialized based on the receiver object type. Similarly, type-directed
cloning~\cite{type_cloning} clones methods based on argument types, which can
produce more specialized code using richer type information. The work of Maxime
Chevalier-Boisvert et al. on Just-In-Time (JIT) specialization for
MATLAB~\cite{mcvm} and similar work done for the MaJIC MATLAB
compiler~\cite{majic_matlab} tries to capture argument types to dynamically
compile optimized copies of functions. All of these techniques are forms of
type-driven code duplication aimed at enhancing type information. Basic block
versioning operates at a lower level of granularity, which allows it to find
optimization opportunities inside of method bodies by duplicating code paths.

\section{Future Work}\label{sec:future}
%
%




Our current implementation only tracks type information intraprocedurally. It
would be desirable to extend basic block versioning in such a way that type
information can cross function call boundaries. This could be accomplished by
allowing functions to have multiple entry point blocks, specialized based on
context information coming from callers. Similarly, call continuation blocks
(return points) could also be versioned to allow information about return types
to flow back into the caller.

Another obvious extension of basic block versioning would be to collect more
detailed type information. For example, we may wish to propagate information
about global variable types, object identity and object field types. It may
also be desirable, in some cases, to know the exact value of some variable or
object field, particularly if this value is likely to remain constant.
Numerical range information could potentially be collected to help eliminate
bound and overflow checks. 

Basic block versioning, as we have implemented it, sometimes generates
versions that account for type combinations that never occur in practice. This
could potentially be addressed by generating stubs for the targets of cloned
conditional branches. Higgs already produces stubs for unexecuted blocks,
but generates all requested versions of a block if the block was ever executed
in the past. Producing stubs for cloned branches would delay the generation of
machine code for these branch targets until we know for a fact that they are
executed, avoiding code generation for unnecessary code paths. The choice of
where to generate stubs could potentially be guided by profiling data.

Some of the information accumulated and propagated by basic block
versioning may not actually be useful for optimization. This is likely to
become a bigger problem if the approach is extended to work accross function
call boundaries, or if more precise type and constant information is accumulated.
An interesting avenue may be to choose which information to propagate based on
usefulness. That is, the most frequently executed type tests are probably the
ones we should focus our resources on. These tests should be dynamically
identified through profiling and used to decide which information to propagate.

\section{Conclusion}\label{sec:conclusion}
We have introduced a novel compilation technique called context-driven
basic block versioning. This technique combines code generation with type
analysis to produce more optimized code through the accumulation of type
information during compilation. The versioning approach is able to perform
optimizations such as automatic hoisting of type tests and efficiently
detangles code paths along which multiple numerical types can occur. Our
experiments show that in most cases, basic block versioning eliminates
significantly more dynamic type tests than is possible using a traditional
flow-based type analysis. It eliminates 64\unskip\% of
type tests on average with a limit of 5 versions per block, compared to
33\unskip\% for the analysis, and never performs worse
than such an analysis.

Basic block versioning trades code size for performance. Such a tradeoff is
often desirable, particularly for performance-critical application kernels.
We have empirically demonstrated that
although our implementation of basic block versioning does increase code
size, the resulting increase is reasonably moderate, and can easily be
limited with techniques as simple as a hard limit on the number of versions
per basic block. In our experiments, a limit of 5 versions per block results
in a mean code size increase of 69\unskip\%.
More sophisticated implementations that adjust the amount of code replication
allowed based on the execution frequency of functions are certainly possible.

Basic block versioning is a simple and practical technique suitable for
integration in real-world compilers. It requires little implementation effort
and can offer important advantages in JIT-compiled environments where type
analysis is often difficult. Dynamic languages, which perform a large number
of dynamic type tests, stand to benefit the most.

Higgs is open source and the code used in preparing this publication is
available on GitHub\footnote{https://github.com/maximecb/Higgs/tree/cc2014}.

\bibliographystyle{splncs03}
\bibliography{paper}

\newpage
\appendix
\section{First Appendix\label{apx:type_prop}}

\begin{algorithm}
\caption{Type propagation analysis}\label{alg:type_prop}
\begin{algorithmic}[1]

\Procedure{typeProp}{$function$}

\State outTypes $\leftarrow \emptyset$
\State edgeTypes $\leftarrow \emptyset$
\State visited $\leftarrow \emptyset$ \Comment{Set of visited control-flow edges}
\State workList $\leftarrow$ $\left\{ \langle null, function.entryBlock \rangle \right\}$

\While{workList not empty}
    \State edge $\leftarrow$ workList.dequeue()
    \State block $\leftarrow$ edge.succ

    \If{block.execCount is 0}
        \State continue \Comment{Ignore yet unexecuted blocks (stubs)}
    \EndIf

    \State visited.add(edge)

    \State $curTypes \leftarrow \emptyset$ \Comment{Merge type info from predecessors}
    \For{edge in block.incoming}
        \If{edge in visited}
            \State curTypes $\leftarrow$ curTypes.merge(edgeTypes.get(edge))
        \EndIf
    \EndFor

    \For{phiNode in block.phis}
        \State t $\leftarrow$ evalPhi(phiNode, block, visited)
        \State curTypes.set(phi, t)
        \State outTypes.set(phi, t)
    \EndFor

    \For{instr in block.instrs}
        \State t $\leftarrow$ evalInstr(instr, curTypes)
        \State curTypes.set(instr, t)
        \State outTypes.set(instr, t)
    \EndFor
\EndWhile

\State \Return outTypes;
\EndProcedure

\algstore{typeprop}
\end{algorithmic}
\end{algorithm}

\begin{algorithm}
\caption{Transfer functions for the type propagation analysis}\label{alg:type_transf}
\begin{algorithmic}[1]
\algrestore{typeprop}

\Procedure{evalPhi}{$phiNode$, $block$, $visited$}
    \State $t \leftarrow \bot$

    \For{edge in block.incoming}
        \If{edge in visited}
            \State predType = getType(edgeTypes.get(edge), phiNode.getArg(edge))
            \If{predType is $\bot$}
                \State \Return $\bot$
            \EndIf
            \State t = t.merge(predType)
        \EndIf
    \EndFor

    \State \Return t
\EndProcedure

\Procedure{AddInt32.evalInstr}{$instr$, $curTypes$}
    \State \Return int32\ \Comment{The output type of AddInt32 is always int32. If an overflow occurs, the result is recomputed using AddFloat64}
\EndProcedure

\Procedure{IsInt32.evalInstr}{$instr$, $curTypes$}
    \State argType $\leftarrow$ getType(curTypes, instr.getArg(0))

    \If{argType is $\bot$} \Comment{If the argument type is not yet evaluated}
        \State \Return $\bot$
    \ElsIf{argType is $\top$} \Comment{If the argument type is unknown}
        \State \Return const
    \ElsIf{argType is int32}
        \State \Return true
    \Else
        \State \Return false
    \EndIf
\EndProcedure

\Procedure{IfTrue.evalInstr}{$instr$, $curTypes$}
    \State arg $\leftarrow$ instr.getArg(0)
    \State argType $\leftarrow$ getType(curTypes, arg)

    \If{argType is $\bot$}
        \State \Return $\bot$
    \EndIf

    \State testVal $\gets$ null
    \State testType $\gets \top$

    \If{arg instanceof IsInt32}
        \State testVal $\leftarrow$ arg.getArg(0) \Comment{Get the SSA value whose type is being tested}
        \State testType $\leftarrow$ int32
    \EndIf

    \If{argType is true or argType is const or argType is $\top$}
        \State queueSucc(instr.trueTarget, typeMap, testVal, testType) \Comment{Queue the true branch, and propagate the test value's type (if applicable)}
    \EndIf

    \If{argType is false or argType is const or argType is $\top$}
        \State queueSucc(instr.falseTarget, typeMap, null, $\top$)
    \EndIf

    \State \Return $\bot$
\EndProcedure

\end{algorithmic}
\end{algorithm}

\end{document}